\documentclass[11pt,a4paper]{article}
\pdfoutput = 1
\usepackage{jcappub}
\usepackage[utf8]{inputenc}
\usepackage{amsmath,amssymb}
\usepackage{graphicx}
\usepackage{hyperref}
\usepackage{natbib}
\usepackage{mathtools}
\usepackage[T1]{fontenc}
\newcommand{\edit}[1]{{#1}}

\newcommand{\lnpk}{2.84 \pm 0.13}
\newcommand{\spk}{0.738 \pm 0.048}
\newcommand{\Ompk}{0.305 \pm 0.01}
\newcommand{\Hpk}{68.5 \pm 1.1}

\newcommand{\lnjoint}{2.81 \pm 0.12}
\newcommand{\sjoint}{0.733 \pm 0.047}
\newcommand{\Omjoint}{0.303 \pm 0.0082}
\newcommand{\Hjoint}{69.23 \pm 0.77}
\newcommand{\SEightjoint}{0.736 \pm 0.051}
\newcommand{\SigEightjoint}{0.735 \pm 0.049}

\newcommand{\bq}{\textbf{q}}
\newcommand{\bx}{\textbf{x}}
\newcommand{\bk}{\textbf{k}}

\newcommand{\bv}{\textbf{v}}

\newcommand{\bs}{\textbf{s}}
\newcommand{\bu}{\textbf{u}}

\newcommand{\half}{\frac{1}{2}}
\newcommand{\third}{\frac{1}{3}}

\def \dtq{\int d^3 \bq \ }

\def\fid{{\rm fid}}
\def\obs{{\rm obs}}
\def\true{{\rm true}}

\newcommand{\avg}[1]{\ensuremath{\left\langle #1 \right\rangle}}

\def\Mpc{\, h^{-1} \, {\rm Mpc}}
\def\Mpccube{\, h^{-3} \, {\rm Mpc}^3}

\def\kMpc{\, h \, {\rm Mpc}^{-1}}

\def\hn{\hat{n}}

\author[a]{Shi-Fan Chen}
\author[b,c,d]{Zvonimir Vlah}
\author[a]{Martin White}

\affiliation[a]{Department of Physics, University of California,
Berkeley, CA 94720.}
\affiliation[b]{Kavli Institute for Cosmology, University of Cambridge, Cambridge CB3 0HA, UK.}
\affiliation[c]{Department of Applied Mathematics and Theoretical Physics, University of Cambridge, Cambridge CB3 0WA, UK.}
\affiliation[d]{Ru\dj er Bo\v skovi\' c Institute, Division of Theoretical Physics, Bijeni\v cka 54, HR-10000 Zagreb, Croatia.}

\emailAdd{shifan\_chen@berkeley.edu}
\emailAdd{zv217@cam.ac.uk}
\emailAdd{mwhite@berkeley.edu}

\title{A new analysis of galaxy 2-point functions in the BOSS survey, including full-shape information and post-reconstruction BAO}


\abstract{We present a new method for consistent, joint analysis of the pre- and post-reconstruction galaxy two-point functions of the BOSS survey.  The post-reconstruction correlation function is used to accurately measure the distance-redshift relation and expansion history, while the pre-reconstruction power spectrum multipoles constrain the broad-band shape and the rate-of-growth of large-scale structure. Our technique uses Lagrangian perturbation theory to self-consistently work at the level of two-point functions, i.e.\ directly with the measured data, without approximating the constraints with summary statistics normalized by the drag scale.  Combining galaxies across the full redshift range and both hemispheres we constrain $\Omega_m=\Omjoint$, $H_0=\Hjoint$ and $\sigma_8=\sjoint$ within the context of $\Lambda$CDM.  These constraints are \edit{consistent} both with the Planck primary CMB anisotropy data and recent cosmic shear surveys. }

\begin{document}
\maketitle
\flushbottom

%------------------------------------------------------------------------------%

\section{Introduction}

The large-scale structure of the Universe, as traced by galaxies, provides fundamental physics information through its connection to both initial conditions in the primordial universe and general relativity through the gravitational formation of structures on the largest observable scales \cite{Pea99,Dod03}. A well-established measure of this structure is the redshift-space two-point function as measured by galaxy redshift surveys \cite{Kaiser87,Ham92}, which encodes both the power spectrum shape of fluctuations in the early universe and, through the quirk that line-of-sight distances in such surveys are inferred from their redshifts, cosmological velocities in the form of redshift-space distortions (RSD).

A particularly interesting and relevant interplay of the initial conditions and gravitational dynamics occurs in the galaxy baryon acoustic oscillations (BAO) signal. The BAO signal is the imprint of early-universe acoustic waves on the observed clustering of galaxies, manifesting as a localized peak in the galaxy correlation function at separations around the characteristic size of these waves \cite{Weinberg13}. The linear physics underlying the size and shape of the BAO feature is well-understood \cite{Eisenstein98,Meiksin99}, making it a robust cosmological signal of both the early universe and the redshift-distance relation. However, the process of nonlinear structure formation leads to a slight wrinkle in this picture: nonlinearities, particular due to bulk displacements of galaxies on large scales, tend to dampen and shift the BAO signal \cite{ESW07,Crocce08}. To better extract the BAO signal, ref.~\cite{ESSS07} proposed a now-standard method known as ``reconstruction'' to cancel a large portion of these effects by estimating large-scale displacements and subtracting them from the observed positions of galaxies.

In recent years there have been significant advances in the modeling and theoretical understanding of both redshift-space distortions and the nonlinear damping of BAO. For the former, the recasting of cosmological perturbation theory in the language of effective field theories has provided a systematic way in which to write down the possible contributions to galaxy clustering on quasilinear scales based on fundamental symmetries while clarifying and taming dependences on small-scale physics (see e.g.\ refs \cite{McDRoy09,BNSZ12,CHS12,Vlah15,Perko16,Vlah18,Chen20,Chen21}). For the latter, improved understanding of the effects of large-scale displacements through so-called infrared (IR) resummations allows us to quantitatively describe nonlinear damping of the BAO peak, both for the raw and reconstructed power spectrum, in perturbation theory \cite{Matsubara08,Carlson13,Vlah16b,Ding18,Chen19b}. These developments allow us to model diverse sets of cosmological observables--- in real and redshift space, power spectrum and correlation function, pre- and post-reconstruction---  within the consistent theoretical framework of perturbation theory.

Our aim in this paper is to present a consistent analysis of the pre- and post-reconstruction 2-point correlation function in Fourier and configuration space using the BOSS survey \cite{Dawson13} as an example. We will work within the framework of Lagrangian perturbation theory \cite{Bernardeau02}, which the present authors have used to develop models for redshift-space distortions and reconstruction \cite{Chen19b,Chen21}. We will operate directly at the level of two-point correlation functions, i.e.\ given a set of cosmological and galaxy bias parameters $\Theta$, we will use the redshift-space power spectra and correlation function multipoles, pre- or post-reconstruction, to construct a likelihood:
\begin{equation}
    \mathcal{L} \propto \exp\Big\{ -\half \big(m(\Theta) - \hat{d}\big)^T C^{-1} \big(m(\Theta) - \hat{d}\big) \Big\} \quad , \quad \hat{d} = (P_\ell, \xi^{\rm recon}_\ell, \cdots ),
\end{equation}
where $m(\Theta)$ is the model, $\hat{d}$ is a data vector composed of two-point correlation functions measured from data and $C$ is the covariance matrix. 

\begin{figure}
    \centering
    \includegraphics[width=\textwidth]{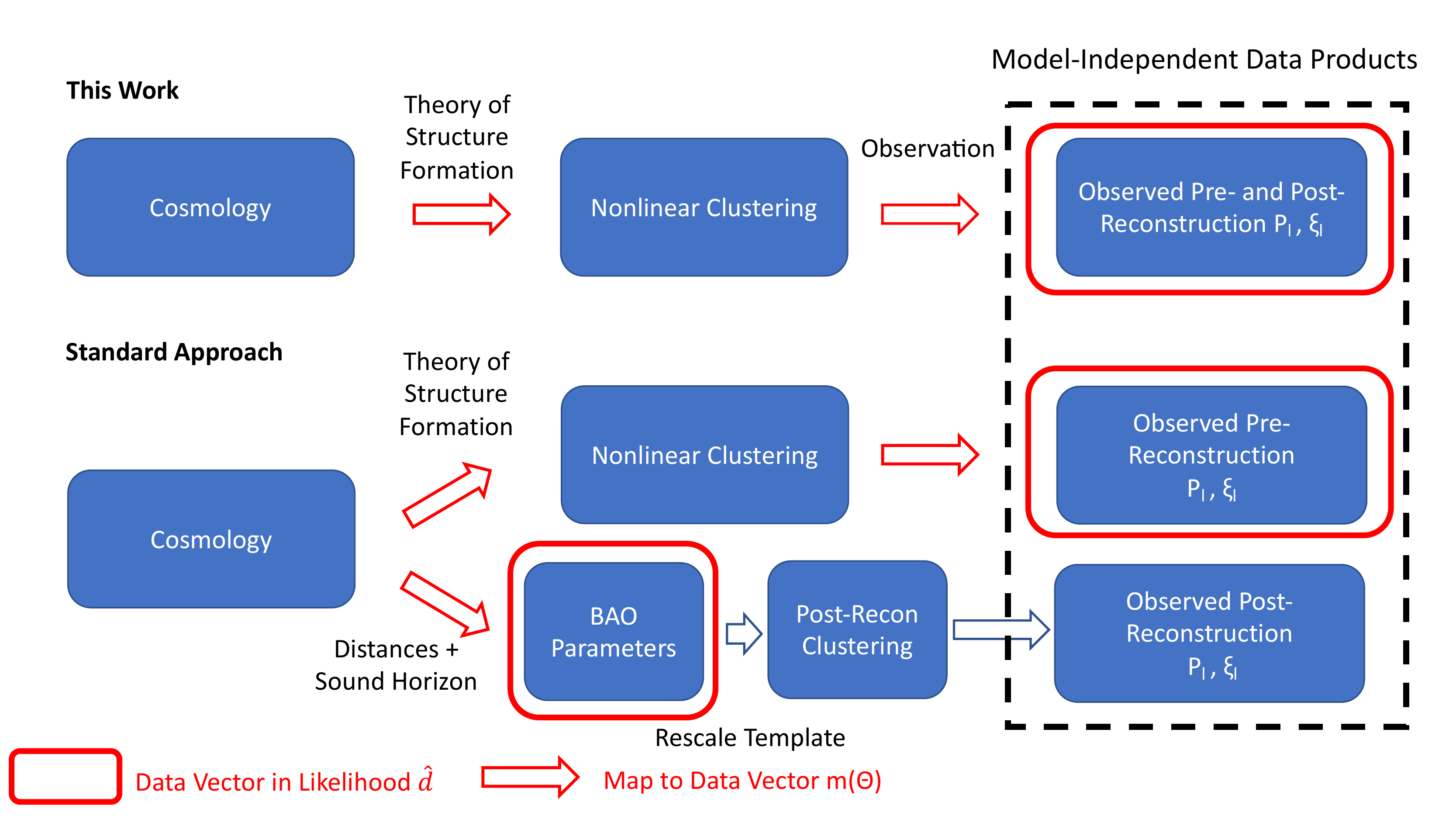}
    \caption{Flowchart comparing our method with the standard approach to combining full-shape RSD and BAO analyses. Arrows denote the unique mapping from theory parameters to observed data. Our approach allows for a direct translation from cosmological parameters into measured 2-point correlation functions \edit{(black dashed box)} via a theory of structure formation (LPT) and does not rely on power spectrum templates or model-dependent BAO parameters derived therefrom. \edit{Squares highlighted in red indicate the actual data vector fit in the likelihoods of each approach, related to cosmological parameters through a model $m(\Theta)$ (red arrows), and each row to the right of ``cosmology'' indicates separate fits which must be combined using simulated mocks.  Our approach features a single fit to the observed data while the standard approach separately fits theory-dependent BAO parameters that depend highly non-linearly on the data and the pre-reconstruction clustering.}
    }
    \label{fig:flowchart}
\end{figure}

This work is not the first to analyze galaxy clustering with an effective theory framework, and follows a number of papers analyzing the BOSS redshift space power spectrum using effective Eulerian perturbation theory \cite{Ivanov20,dAmico20} as well as work by refs.~\cite{Philcox20,DAmico21} combining these analyses with additional BAO information through reconstruction. Our goal, rather, is to perform a joint analysis of pre- and post-reconstruction data without resorting to additional assumptions or machinery \edit{(Fig.~\ref{fig:flowchart})}. In particular, previous work combining pre- and post-reconstruction measurements (e.g.\ refs~\cite{Beutler17,Philcox20,DAmico21}) have typically sought to distill the content of the latter by fitting a set of BAO parameters $\tilde{\alpha}_{\parallel,\perp}$ (\S~\ref{ssec:bao}) which scale the BAO signal in a template, or fiducial, linear power spectrum to match that of the observed signal. Then, since the BAO oscillations in the template differ from the observed by the cosmological dependence of the sound horizon ($r_d$) in addition to simple distance scalings by redshift, these \edit{best-fit} $\tilde{\alpha}$'s are fit to the ratio of cosmological distances to $r_d$ in conjunction to the pre-reconstruction power spectra\edit{, such that the data vector is instead $\hat{d} = (P_\ell, \tilde{\alpha}_\parallel, \tilde{\alpha}_\perp)$}. The covariance of the power spectra and $\tilde{\alpha}$'s is then inferred from measurements of $\tilde{\alpha}$ from the power spectra in mock catalogs. \edit{A schematic comparing our approach in this work to the standard one is shown in Fig.~\ref{fig:flowchart}. F}or a given set of cosmological parameters the information about $r_d$ and cosmological distances is inherent in the perturbation-theory prediction $\xi^{\rm recon}_\ell(\Theta)$, allowing us to bypass the need to approximate the BAO signal as a scaled version of a fixed power spectrum template and directly compare cosmology with data. In particular, since the BAO information in the correlation function is effectively isolated in configuration space as a sharp peak at large scales, our fiducial setup will combine the post-reconstruction correlation function around the peak with a full-shape analysis of the pre-reconstruction power spectrum. \edit{Note that while we have focused our discussion on the post-reconstruction BAO measurement, the BOSS collaboration (but not refs.~\cite{Philcox20,DAmico21}) similarly also distilled the RSD signal pre-reconstruction into a best-fit $f\sigma_8$, such that the final data vector fit to cosmological parameters was $\hat{d} = (f\sigma_8, \tilde{\alpha}_\parallel, \tilde{\alpha}_\perp$).}

\edit{The simplified approach advocated for in this work has a number of advantages for obtaining cosmological constraints from surveys like BOSS. For a given set of theory parameters (including cosmology and galaxy bias) there is a unique `forward' mapping --- shown in the top row of Fig.~\ref{fig:flowchart} --- from these parameters to the (model-independent) observational data. This implies that constraining information in the data that is implied by the theory model is captured without loss. By comparison, in the standard approach one assumes e.g.\ that any BAO information post-reconstruction can be distilled into two BAO scaling parameters which, by themselves, do not uniquely map into the space of post-reconstruction observables, potentially leading to information loss (\S\ref{sec:model}). Furthermore, while the standard approach requires measuring the covariance of these summary statistics from approximate mock catalogs, computing the likelihood at the level of model-independent data in principle allows us to straightforwardly use the statistical uncertainties implied by the theory model itself, e.g.\ by computing covariance matrices analytically within perturbation theory \cite{Wadekar20}\footnote{See, however, ref.~\cite{Philcox20} for an idealized calculation using Fisher matrices.}. Future theory calculations of noise-free data covariances have the potential to ease numerical difficulties from estimating covariances of statistics beyond the pre-reconstruction power spectrum with only a finite number of mocks.}

The outline of the paper is as follows.  In \S\ref{sec:data} we present the galaxy samples that we analyze, which are all drawn from the BOSS survey \cite{Dawson13}.  The models we fit to these data, all based on cosmological perturbation theory, are described in \S\ref{sec:model}.  Our fiducial analysis setup, including scale cuts, parameter choices and priors are presented in \S\ref{sec:setup}. Our final cosmoloical constraints are given in \S\ref{sec:results}, where we also discuss constraints from different subsamples of the BOSS data and compare to constraints from other groups and experiments. We conclude in \S\ref{sec:conclusions}.
Some technical details are relegated to a series of Appendices. 

\section{Data}
\label{sec:data}

\begin{figure}
    \centering
    \includegraphics[width=\textwidth]{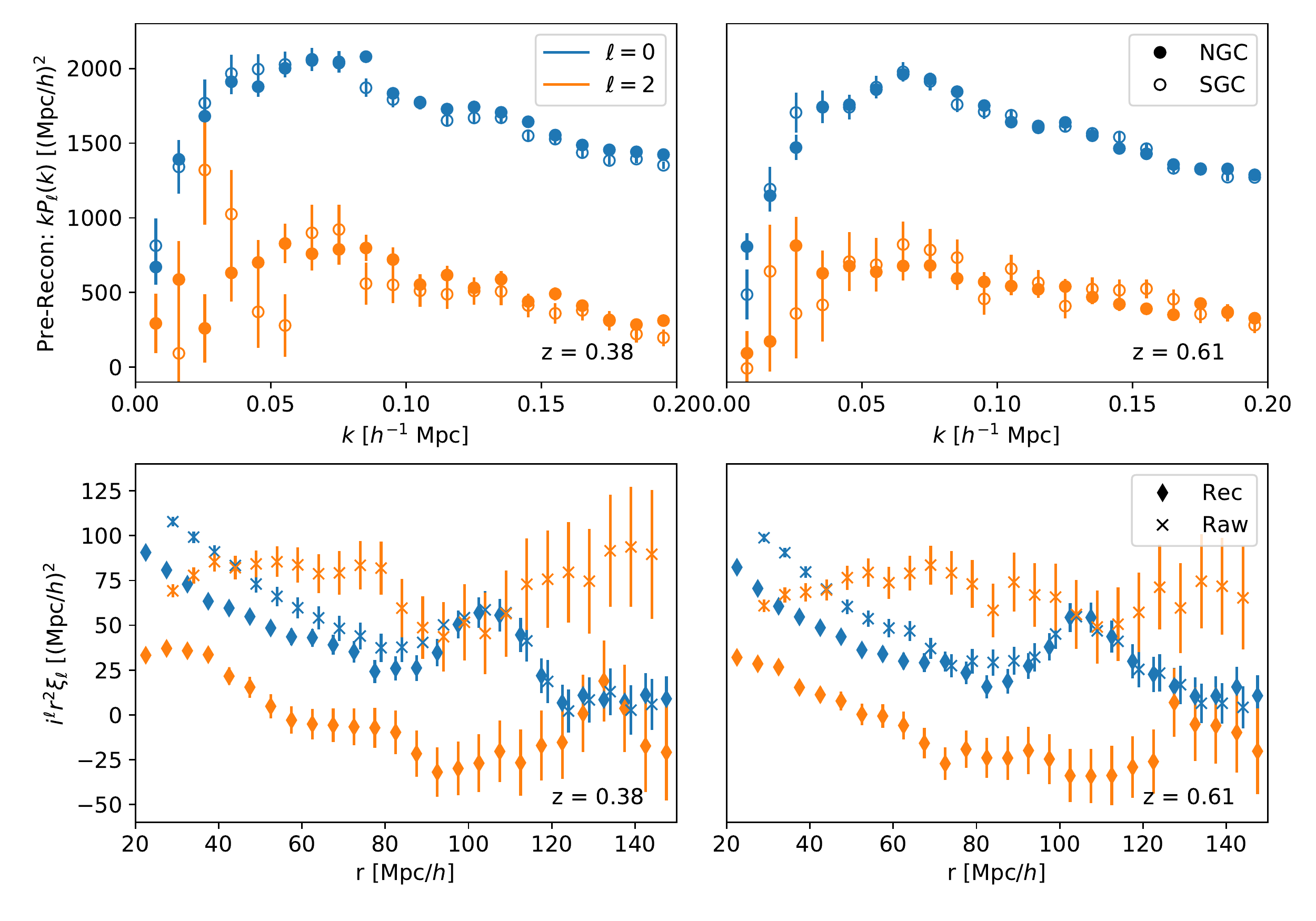}
    \caption{The pre-reconstruction power spectrum (top) and pre- and post-reconstruction correlation functions (bottom) of the BOSS DR12 galaxies. The monopole and quadrupole of each are both shown in blue and orange, respectively. For the power spectrum, the separate measurements for the NGC and SGC samples are shown as \edit{filled and open circles}. Pre- and post-reconstruction (``Raw'' and ``Rec'') correlation function measurements are shown with round and crossed markers. The correlation function is measured jointly across both galactic caps. Error bars represent the diagonals of the covariance matrix computed using 1000 Patchy mocks.
    }
    \label{fig:data}
\end{figure}

We analyze the clustering of galaxies drawn from the BOSS galaxy redshift survey \cite{Dawson13}, part of the Sloan Digital Sky Survey III \cite{SDSSIII}. Galaxies in BOSS were targeted with two independent selection criteria: one (LOWZ) targeted luminous red galaxies up to $z = 0.4$ while another (CMASS) targeted massive galaxies with $0.4 < z < 0.7$; however, due to an incorrect application of the LOWZ criteria in the first nine months of the survey, two additional samples (LOWZE2, LOWZE3) had to be separated out. All of these samples are described in more detail in ref.~\cite{Reid16}. Ref.~\cite{Alam17} combined these samples into three redshift bins with $0.2<z<0.5$, $0.4<z<0.6$ and $0.5<z<0.75$ (named \textbf{z1}, \textbf{z2} and \textbf{z3}, respectively). Since \textbf{z2} overlaps both \textbf{z1} or \textbf{z3} in redshift and thus gives correlated constraints we choose to analyze \textbf{z1} and \textbf{z3} only in this work. Each redshift bin can be further split into galaxies observed in the Northern (NGC) and Southern (SGC) galactic caps.  Because the imaging in the north and the south differ slightly, the samples have slightly different properties and should be analyzed separately.  The final BOSS sample covers 1,198,006 galaxies in total over 10,252 square degrees of sky.

The power spectra and correlation function multipoles of these samples were measured in refs.~\cite{Beutler17,Vargas18} and the data are shown in Fig.~\ref{fig:data}. The BOSS two-point function measurements were computed assuming a flat $\Lambda$CDM cosmology with present-day matter density $\Omega_{M, \rm fid} = 0.31$. This implies that the reported redshift-space power spectrum is related to its value in the coordinates of the true cosmology by
\begin{equation*}
    P^{\rm obs}_s(\bk_{\rm obs}) = \alpha_\perp^{-2} \alpha_\parallel^{-1} P_s(\bk)
    \quad , \quad
    k^{\rm obs}_{\parallel, \perp} =  \alpha_{\parallel, \perp}  k_{\parallel, \perp}
    \quad ,
\end{equation*}
where the Alcock-Paczynski parameters are defined as \cite{Alcock79,Padmanabhan08}
\begin{equation}
    \alpha_\parallel = \frac{H^{\rm fid}(z)}{H(z)}
    \quad , \quad
    \alpha_\perp = \frac{D_A(z)}{D^{\rm fid}_A(z)} \quad .
\label{eqn:AP}
\end{equation}
Since all distances are reported in $h^{-1}$Mpc units the above ratios should be computed assuming a fixed $h$. The equivalent relations for the correlation function are simply the Fourier transforms of the above equations. The mismatch between true and fiducial coordinates imprints additional anisotropy in the galaxy two-point function and serves as a further source of cosmological information known as the Alcock-Paczynski (AP) effect \cite{Alcock79} (Appendix~\ref{app:template_fit}).

For the power spectrum we use the updated pre-reconstruction measurements of both data and mock catalogs presented in ref.~\cite{Beutler21}. On top of the AP effect the geometry of the survey itself leaves an imprint in the clustering of the galaxies, so that the measured power spectra are the convolution of the true clustering signal with a window function and taking into account wide-angle effects \cite{FKP,Wilson17,Castorina18,Beutler19}; to this end ref.~\cite{Beutler21} provide for each sample a wide angle matrix $\textbf{M}$ and window function matrix $\textbf{W}$ such that for an input vector of appropriately-binned theoretical (including AP) power spectrum multipoles $\textbf{P}$ the observed (binned) power spectra are given by\footnote{\edit{We use the updated values of the power spectrum multipoles and window functions from a revised version of ref.~\cite{Beutler21}. These measurements fix a mismatch in the normalizations of the window function and power spectra at the roughly $10\%$ level that afflicted earlier results and which improves the agreement with correlation function fits which do not require multiplying by a window function. This normalization issue is discussed in detail in the published version of ref.~\cite{Beutler21}, with resulting amplitude corrections to each BOSS sample tabulated in Table 1 of that work; roughly, there is an overall degeneracy between the power spectrum and window function amplitudes, such that no results are affected when both are adjusted simultaneously by a single multiplicative factor--however, this is premised upon the two being normalized consistently when computed. We thank Pat McDonald and Florian Beutler for helpful discussions of this issue. Upated versions of the BOSS power spectra and window functions can be found in \url{https://fbeutler.github.io/hub/deconv_paper.html}.}}
\begin{equation*}
    \textbf{P}^{\rm conv} = \textbf{W}\textbf{M} \textbf{P}, \quad \textbf{P} = (P_0, P_2, P_4).
\end{equation*}
While the output $\textbf{P}^{\rm conv}$ contains both the hexadecapole as well as odd multipoles $P_{1,3}$ we will restrict our analysis in this paper to the monopole and quadrupole only.

For our post-reconstruction BAO analysis we use the post-reconstruction correlation function multipoles obtained in ref.~\cite{Vargas18}. Both pre- and post-reconstruction correlation function multipoles were measured for the combined NGC and SGC samples at the redshift bins \textbf{z1} and \textbf{z3}. The reconstruction \edit{in the public BOSS data} was performed using the so-called \textbf{RecIso} convention, with reconstructed displacements solved-for using a finite-difference approach on the observed galaxy density field smoothed by a Gaussian filter with width $R = 15 h^{-1}$ Mpc. We will discuss further details of the procedure in Section~\ref{ssec:bao}.

\begin{figure}
    \centering
    \includegraphics[width=\textwidth]{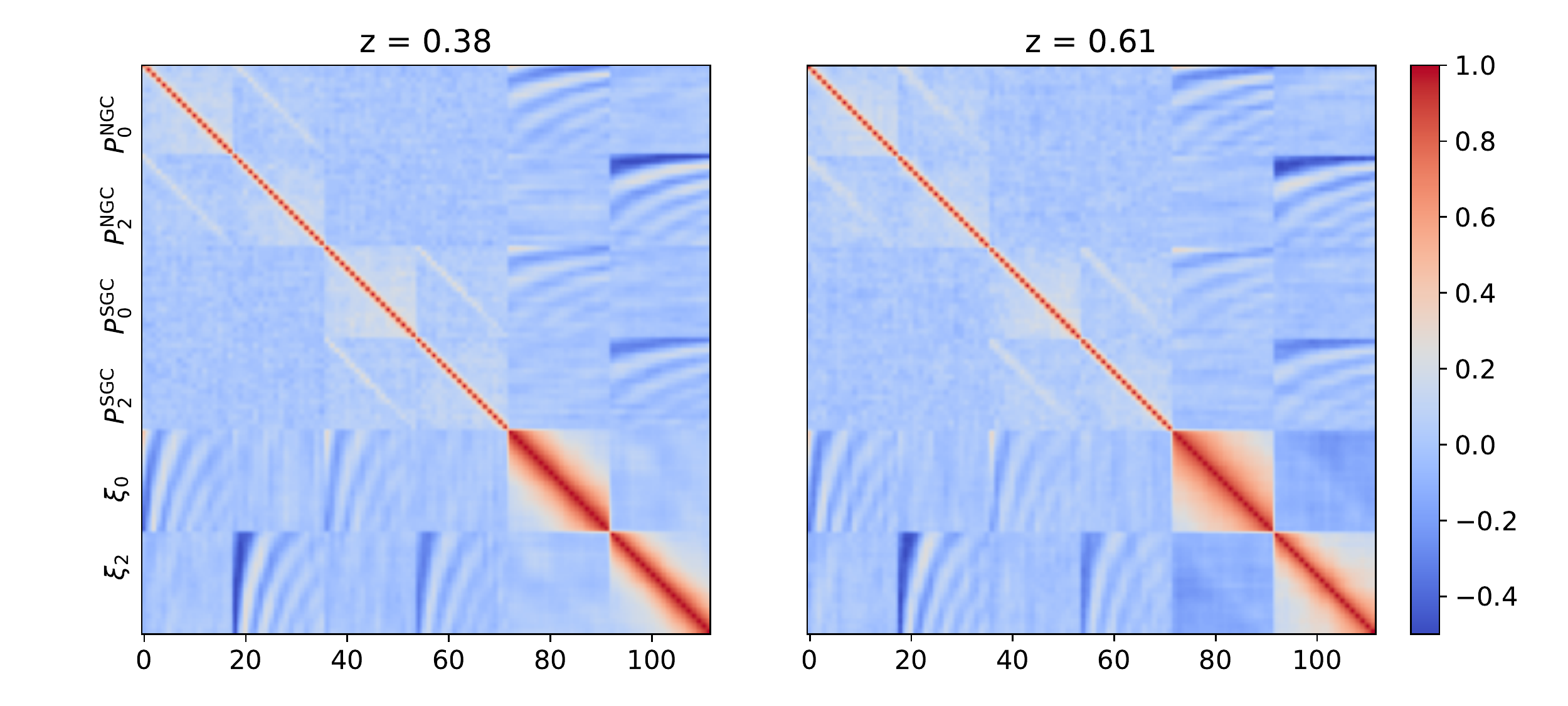}
    \caption{The joint correlation matrix of the \textbf{z1} and \textbf{z3} pre- and post-reconstruction two-point function samples, computed using 1000 Patchy mocks. For clarity of presentation we have restricted the power spectrum to wavenumbers $0.02 \kMpc < k  < 0.20 \kMpc$ (18 bins) and the correlation function to $80\,h^{-1}\text{Mpc} < r < 130\, h^{-1}\text{Mpc}$ (20 bins). The numbers on the $x$ axis denote bin number.  }
    \label{fig:cov}
\end{figure}

Finally, to obtain the joint covariances of the power spectrum\footnote{For the power spectrum mocks, see: \url{https://fbeutler.github.io/hub/deconv_paper.html}.} and correlation function\footnote{We thank Mariana Vargas for providing the correlation function mock measurements.} measurements used in our analysis we used the V6C BigMultiDark Patchy mocks \cite{Kitaura16} released with DR12 of the SDSS-III Survey. These mocks were prepared using approximate gravity solvers with galaxy biasing calibrated to the BigMultiDark simulation in a redshift-dependent way to capture the time-depedence of the BOSS galaxy sample \cite{Kitaura16}. The thus-derived correlation matrix for the power spectrum and post-reconstruction correlation function is shown in Fig.~\ref{fig:cov}. Since the correlation-function mock measurements were only obtained for 1000 of these mocks that is also the number of power spectrum measurments we use to obtain the joint pre- and post-reconstruction covariance. \edit{The thousand mocks are sufficient for our purposes; concretely, our most extensive analysis setup will include (for each independent redshift bin) $18$ $k$-bins per power spectrum multipole per galactic cap and $10$ radial bins per correlation function multipole, yielding a $92$-element data vector.  Applying a multiplicative correction to unbias the precision matrix \cite{Hartlap2007} would lead to rescalings of the parameter errors of less than $5\%$, with no change in the matrix structure.} As a further test, reducing to $N_{\rm mocks} = 500$ changed the $\chi^2$ of the best-fit model found for the \textbf{z3} sample by $\Delta \chi^2 = 3.5$, i.e.\ far less than one per d.o.f.\ for our fiducial setup (\S~\ref{ssec:scale_cuts}), suggesting that one thousand mocks measurements is sufficient for our purposes. Finally, while we have not found it necessary in our analysis, we note that if we were to rebin the power spectrum data into broader $k$ bins we would reduce the number of degrees of freedom and more cleanly separate the BAO and broadband shape information between the correlation function and power spectrum. This may be beneficial in future analyses. 

\section{Model}
\label{sec:model}

Our aim in this work is to jointly model the redshift-space galaxy two-point function both pre- and post-reconstruction within a consistent theoretical framework. Specifically, we will operate within Lagrangian perturbation theory (LPT; \cite{Bernardeau02,Matsubara08}), which models nonlinear structure formation through the evolution of galaxy displacements $\Psi(\bq,\tau)$, relating observed (Eulerian) positions $\bx$ at a given time $\tau$ to initial (Lagrangian) positions $\bq$ through $\bx = \bq + \Psi$. The displacements are expanded order-by-order in the initial conditions $\Psi = \Psi^{(1)} + \Psi^{(2)} + \Psi^{(3)} + ...$. In this work we will operate within the EdS approximation wherein the n$^{\rm th}$ order displacement scales as the n$^{\rm th}$ power of the linear growth factor $D(z)$; this has been shown to be an excellent approximation for current and upcoming galaxy surveys in \cite{Takahashi:2008, Fasiello+:2016, delaBella:2017, Fujita+:2020, Donath+:2020}, with a slight caveat due to scale dependence from massless neutrinos which we will address in Section~\ref{sec:setup}. The nature of the mapping between Lagrangian and Eulerian coordinates makes LPT a natural arena in which to understand both redshift-space distortions and nonlinear BAO damping, which we discuss in term below. In addition, in order to speed up our calculations and avoid repeatedly calling Boltzmann codes at each new point in our Markov chains, in this paper we have chosen to approximate our theory components as a Taylor series in the cosmological parameters. The grid of PT predictions used to compute the Taylor-series coefficients was computed using \texttt{CLASS} \cite{CLASS} and \texttt{velocileptors} \cite{Chen20,Chen21}.  The details of our approximation scheme are discussed in Appendix~\ref{app:taylor}.

\subsection{One-Loop Redshift Space Power Spectrum}
\label{ssec:full_shape}
In spectroscopic galaxy surveys like BOSS a galaxy's position along the line of sight (LOS) is inferred from its measured redshift. Since a galaxy's cosmological redshift and peculiar velocities both contribute to this redshift, its redshift-space position $\bs$ is boosted along the LOS by its peculiar LOS velocity in the appropriate units, that is $\bs = \bx + \bu$, where $\bu = (\hn \cdot \bv) \hn/ \mathcal{H}$ and $\mathcal{H}$ is the conformal Hubble parameter \cite{Kaiser87}. Within LPT this is equivalent to boosting the displacement $\Psi$ by the LOS component of its (appropriately normalized) time derivative, i.e. $\Psi_s = \Psi + \dot{\Psi}$ \cite{Matsubara08,Chen21}; conveniently, this boost can be recast as a coordinate transformation
\begin{equation}
    \Psi^{(n)}_{s,i} = \Psi^{(n)}_i + n f \hn_i \hn_j \Psi^{(n)}_j \equiv R^{(n)}_{ij} \Psi^{(n)}_j \quad .
\end{equation}
For the remainder of this paper we will in addition make the plane-parallel approximation that the LOS vector $\hn$ is a constant independent of position.

The translation between densities and displacements follows from number conservation. Assuming galaxies are sampled from the initial conditions $\delta_0$ according to some functional $\rho_g(\bq) = F[\delta_0(\bq)]$ this implies that the present day density satisfies $\rho_g(\bx) d^3\bx = \rho_g(\bq) d^3\bq$ or, in Fourier space, \cite{Matsubara08,Vlah16,Desjacques18,Chen21}
\begin{equation}
    1 + \delta_g(\bk) = \dtq e^{i\bk\cdot(\bq+\Psi)} F(\bq) \quad .
\end{equation}
The equivalent equation in redshift space follows by substituting $\Psi$ for $\Psi_s$. The bias functional $F(\bq)$ can be perturbatively expanded as
\begin{equation}
    F(\bq) = b_1 \delta_0(\bq) + \frac{1}{2} b_2 (\delta_0(\bq)^2 - \avg{\delta_0^2}) + b_s (s^2_0(\bq) - \avg{s^2})
\end{equation}
where $s^2_0 = (\partial_i \partial_j / \partial^2 - \delta_{ij}/3) \delta_0$ is the shear tensor. For a complete accounting of terms relevant to the one-loop power spectrum we in principle have to also account for third order bias operators; however, since we expect them to be small for all but the most massive halos \cite{Fujita16,Ivanov20,Chen20}, and highly degenerate with the effective-theory corrections discussed below, we will set them to zero for the rest of this work.

The power spectrum in redshift space is then given by \cite{Matsubara08,Chen21}
\begin{equation}
    P_s(\bk) = \dtq \avg{ e^{i\bk\cdot(\bq+\Delta_s)} F(\bq_1) F(\bq_2)}_{\bq = \bq_1-\bq_2},
    \label{eqn:fseqn}
\end{equation}
where we have defined the pairwise displacement in redshift space $\Delta_s = \Psi_s(\bq_1) - \Psi_s(\bq_2)$. In the case of matter ($F=1$) within first-order LPT (Zeldovich approximatin) Equation~\ref{eqn:fseqn} can be evaluated exactly to give $P_{\rm Zel} = \int d^3\bq \exp [i k_i q_i - k_i k_j A_{ij}/2 ]$ where we have defined the second cumulant of pairwise displacements $A_{ij} = \avg{\Delta_i \Delta_j}$. The exponentiation of the pairwise displacement in Equation~\ref{eqn:fseqn} is highly significant and allows LPT to capture the nonlinear damping of BAO due to the large-scale (bulk) displacements \cite{Senatore14,Vlah15,Baldauf15,Vlah16,Blas16}. In practice we keep only these long-wavelength displacements exponentiated and perturbatively expand those above a certain wavenumber ($k_{\rm IR} = 0.2 \kMpc$) perturbatively; for further details we refer interested readers to ref.~\cite{Chen21}, which also provides a detailed exposition of the various terms implied in Equation~\ref{eqn:fseqn} as well as relevant numerical methods. Finally, as an effective perturbation theory LPT requires a number of counterterms to properly tame its sensitivity to small-scale (UV) physics; in this work we adopt the parametrization of ref.~\cite{Chen20} and write
\begin{equation}
    P_s(\bk) = P_s^{\rm PT}(\bk) + (\alpha_0 + \alpha_2 \mu^2) k^2 P_{\rm Zel}(\bk) + R_h^3 ( 1 + \sigma^2 k^2 \mu^2 )
\end{equation}
where $P_{\rm Zel}$ is the Zeldovich matter power spectrum. Note that while the above parametrization is not exhaustive (i.e.\ we should in principle include terms like $\alpha_4 \mu^4$), this parameter set has been tested extensively against simulations (e.g.\ refs.~\cite{Chen19b,Chen20,Chen20b,Chen21}) and the neglected terms are extremeley degenerate with those listed when fitting the monopole and quadrupole only. The above parameter set can also be used to fit the pre-reconstruction correlation function, which can be obtained directly by Fourier transforming the theory prediction for the power spectrum.

\subsection{Nonlinear BAO Damping Post Reconstruction}
\label{ssec:bao}

It is well known that nonlinear structure formation smooths out the BAO peak in the galaxy two-point function \cite{Bharadwaj96,ESW07,Matsubara08,Crocce08,Padmanabhan09,Noh09}, reducing its prominence and as a result also the signal-to-noise of BAO measurements. Within LPT this phenomenon can be understood by looking at contributions to Equation~\ref{eqn:fseqn} such as \cite{Chen20b}
\begin{equation*}
    P_{\rm BAO}(\bk) \sim \dtq e^{i\bk \cdot \bq - \half k_i k_j A_{ij}(\bq)}\ \xi_{\rm BAO}(\bq) \approx e^{-\half k^2 \Sigma^2} \dtq e^{i\bk \cdot \bq }\ \xi_{\rm BAO}(\bq).
\end{equation*}
Since the BAO peak is well-localized at the sound horizon at the drag epoch $r_d$, it acts to pick out a particular scale at which to evaluate $A_{ij}$, from which a specific damping scale $\Sigma^2 = \avg{A_{ij}}_{|\bq|=r_d}$, leading to Gaussian damping of the BAO. Within $\Lambda$CDM most of the displacement power comes from relatively low wavenumbers \cite{ESW07}.

The purpose of standard reconstruction \cite{ESSS07} is to sharpen the BAO feature by undoing some of this damping. To do so, one smooths the observed galaxy density field using a Gaussian filter $S(k)$ on a sufficiently large scale that the Kaiser formula $\delta_g(\bk) = (b + f\mu^2) \delta_m$ is a good approximation, uses the smoothed field to solve for the (smoothed) linear Zeldovich displacement $S(k) \Psi_{\rm Zel}(\bk) = i \bk /k^2 S(k) \delta_m$, and subtracts these displacements from the observed galaxy positions. To preserve power on large scales a random catalog is shifted using the same displacements, with the difference between the displaced galaxies ($d$) and shifted randoms ($s$) constituting the full reconstructed density field $\delta_{\rm rec} = \delta_d - \delta_s$. The presence of redshift space distortions presents a slight complication and there is no general settled-upon convention in the literature: the BOSS data analyzed in this work followed the so-called \textbf{RecIso} convention wherein the galaxies were displaced with RSD (that is, the reconstructed displacement multiplied by $R^{(1)}_{ij}$) while the randoms were shifted without RSD \cite{Padmanabhan12}.

The residual damping left in the BAO signal after reconstruction can be modeled in the same way as was the damping pre-reconstruction. Specifically, we can think of the displaced galaxies ($d$) and shifted randoms ($s$) as two additional tracers with displacements \cite{Padmanabhan09,Noh09,White15,Chen19b}
\begin{equation}
    \Psi^d_i = R^{(1)}_{ij} (1 - \mathcal{S}) \ast \Psi^{\rm Zel}_j
    \quad , \quad
    \Psi^s_i = - \mathcal{S} \ast \Psi^{\rm Zel}_i.
\end{equation}
The two-point statistics of these displacements can then be computed at the sound horizon $r_d$ as in the pre-reconstruction case and used to damp the BAO feature. Specifically, for a given linear power spectrum $P_{\rm lin}$ we can decompose it into a ``wiggle'' component with the BAO feaure and a smooth component without, i.e.\ $P_{\rm lin} = P_w + P_{nw}$. The prediction for each (cross) spectrum between $d$, $s$ is then
\begin{equation}
    P^{ab}(\bk) = K^{ab}(k,\mu) \left[ P_w(k) e^{-k^2 \Sigma^{ab}(\mu)/2} + P_{nw}(k) \right]
\label{eqn:Pab}
\end{equation}
where $K^{ab}(k,\mu)$ is a linear-theory function of the growth rate $f(z)$ and linear bias $b_1$ taking into account Kaiser infall and the smoothing filter. 

\edit{Equation~\ref{eqn:Pab} captures (resums) the nonperturbative effect of bulk displacements on linear-theory BAO wiggles; in principle, nonlinear effects such as mode coupling can induce further modifications to the BAO feature. A full treatment of these additional effects for galaxies at one-loop order is complex and beyond the scope of this work (see however ref.~\cite{Hikage20} for the calculation in the case of matter only).  Calculations taking into account nonlinear bias within the Zeldovich approximation show that phase shifts due to mode coupling are substantially reduced post-reconstruction, suggesting that the in-phase damping of the BAO feature due to IR displacements is the dominant nonlinear effect \cite{Chen19b}. However, in order to further} insulate our full-shape fits from potential systematics of the reconstruction procedure\edit{, including due to residual nonlinear contributions,} we set the linear bias and growth rate in the BAO model to be free parameters called $F$ and $B_1$, and in addition allow for broadband deviations between the theory predictions for $\xi_\ell$ and the data by fitting adding a linear template to the theory predictions, that is
\begin{equation}
    \xi_\ell(r) = \xi^{\rm th}_\ell(r) + a_{\ell,0} + \frac{a_{\ell,1}}{r}
\end{equation}
where $\xi^{\rm th}_\ell$ are the multipoles of the Fourier transformed (with AP) theory predictions of Equation~\ref{eqn:Pab}. Detailed functional forms and integrals are given in Appendix~\ref{app:sigma_bao}.

Let us conclude this section by comparing our approach in BAO fitting with that in previous work. Traditionally, the BAO have been fit using a template power spectrum whose BAO component is scaled by
\begin{equation}
    \tilde{\alpha}_{\parallel,\perp} = \Bigg(\frac{r^\fid_d}{r_d}\Bigg)\ \alpha_{\parallel,\perp}.
\end{equation}
This takes into account stretching of the BAO signal due both to changing distance scales ($\alpha_{\parallel,\perp}$) and sound horizon ($r_d$). Typically this template model is used to fit directly for the $\tilde{\alpha}$'s, whose output likelihoods can then be used in broader cosmological analyses. Note that this prescription mixes two in-principle distinct physical effects: the cosmological dependence of the sound horizon $r_d$ and the anisotropy due to the Alcock-Paczynski effect ($\alpha_{\parallel,\perp}$), While the latter would exist even in the absence of the BAO peak, the conventional method assumes the BAO carries the bulk of the signal, with other effects from the broadband shape of the power spectrum discarded, or marginalized away, via polynomial parameters.

In this work we eschew the use of templates for both pre- and post-reconstruction fitting, choosing instead to extract the BAO wiggles for each cosmology directly from the transfer-function outputs of Boltzmann codes, with the perspective that Equation~\ref{eqn:fseqn} and \ref{eqn:Pab} are definite predictions of the pre- and post-reconstruction galaxy two-point function for any given cosmology. This obviates the need for the $r_d$ scaling in $\tilde{\alpha}$, since the frequency of the BAO wiggles is automatically encoded in the linear power spectrum, and automatically includes any distance and anisotropy information present in the data. \edit{Modeling the reconstructed BAO at the data instead of the summary statistic level also has advantages for constraining non-standard physical effects like beyond $\Lambda$CDM physics. While measurements of the BAO scale are robust to many such physical effects \cite{Bernal20}, there are notable and theoretically well-understood exceptions, for example features around the BAO scale due to relative perturbations between baryons and dark matter \cite{Schmidt17,Chen19}. Beyond robustness, measurements of BAO scaling parameters alone cannot capture many effects related to oscillatory features in the power spectrum that can be sharpened by reconstruction, including neutrino or light-relic induced phase shifts in the BAO \cite{Baumann18} or inflationary signatures \cite{Vasudevan19,Beutler19b,Chen20b}. Indeed, ref.~\cite{Baumann20} detected the neutrino-induced phase shift in reconstructed BOSS data by expanding the standard template fit with an additional phase-shift parameter. On the other hand, for a given cosmological model, any such effects are automatically included when fitting at the data level without modification and, once included in a theory model, cannot act as a theoretical systematic by default. F}or completeness, and to correct some typos in the literature, we include a detailed description of the standard method in Appendix~\ref{app:template_fit}.

\section{Analysis Setup}
\label{sec:setup}

In this paper we aim to perform a joint analysis combining full-shape information in pre-reconstruction power spectrum and additional information in the post-reconstruction correlation function. To this end, our fiducial setup will combine the former with the latter in a narrow band around the peak where most of the BAO information is isolated. Our goal in this section is to outline and explain this setup, and lay out the accompanying cosmological and effective-theory parameter choices and priors.

\subsection{Scale Cuts}
\label{ssec:scale_cuts}

\begin{figure}
    \centering
    \includegraphics[width=\textwidth]{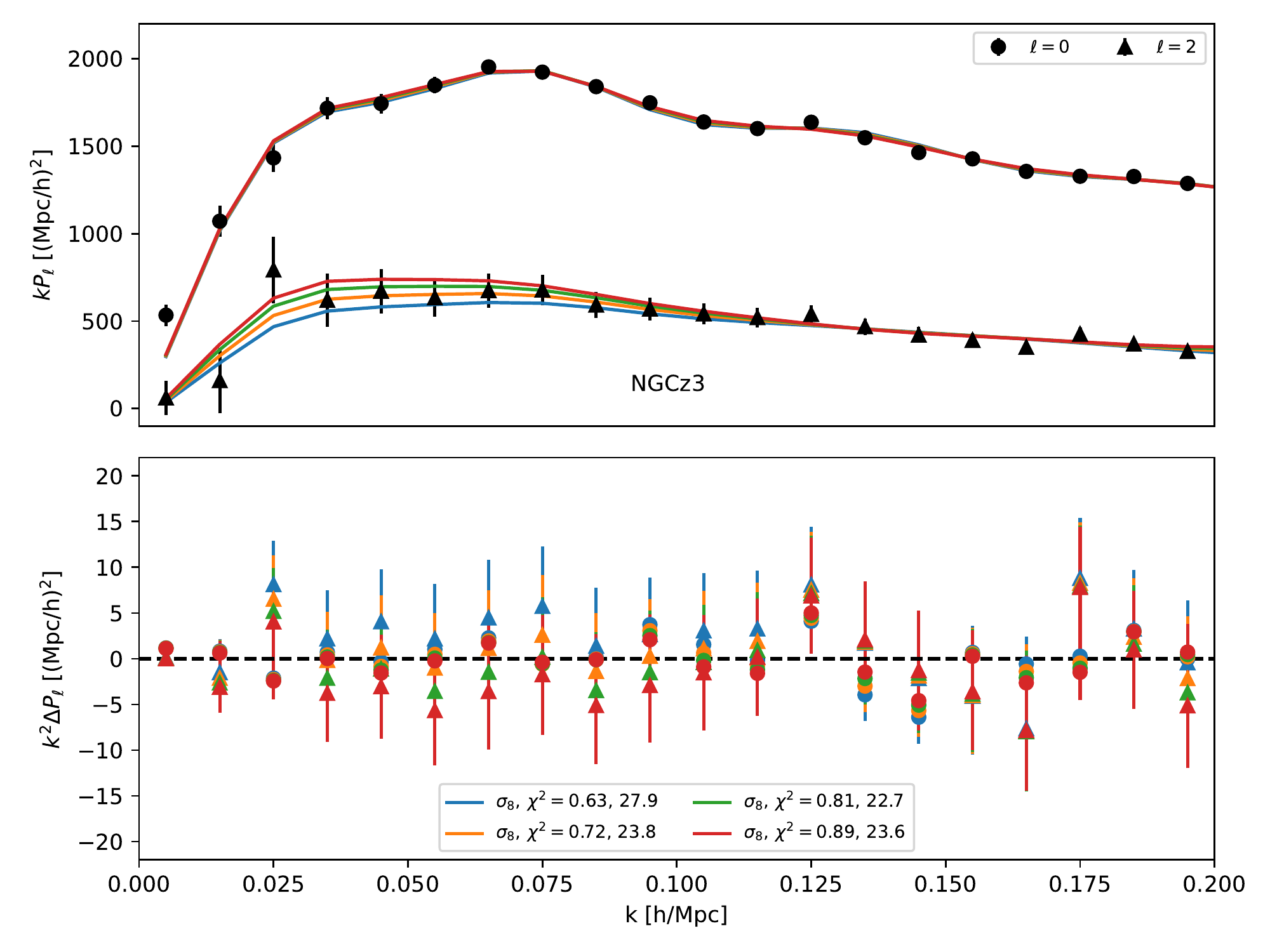}
    \caption{Best fit models to the \textbf{NGCz3} pre-reconstruction power spectrum multipoles (top) and their residuals (bottom), fixing $\Omega_m$ and $h$ to their Planck best-fit values and a scan in $\sigma_8$ (different curves; see legend).  The biases and counterterms are varied to find the best fit for each $\sigma_8$ and so differ between curves.  The models primarily vary in their predictions for the low $k$ quadrupole amplitude while making essentially identical predictions to the high $k$ amplitude across different $\sigma_8$'s.  This suggests the constraining power on $\sigma_8$ comes primarily from large scales.
    }
    \label{fig:sig8_scan}
\end{figure}

We begin by setting up the Fourier-space side of our analysis. Throughout this paper we will adopt the Fourier-space scale cuts $k_{\rm min} = 0.02 \kMpc$ and $k_{\rm max}=0.20 \kMpc$. The LPT model we use in this work was shown to yield unbiased cosmological constraints on this range of scales for BOSS-like samples even at significantly larger volumes \cite{Chen21}. We drop the lowest two $k$ bins below $0.02 \kMpc$; including them increased the best-fit $\chi^2$ for the \textbf{NGCz3} sample when limiting to a Planck cosmology by $\Delta \chi^2 = 20$, suggesting systematic errors in the data beyond our theoretical modeling. 

It is worth noting that we have chosen to adopt a more conservative scale cut than some other works in the literature, like Ref.~\cite{Ivanov20} who adopt $k_{\rm max} = 0.25 \kMpc$, even though we expect our theoretical model to perform as well as other effective-theory models on the pre-reconstruction power spectrum. We have made this choice on the reasoning that the information on the primordial (linear) power spectrum rests primarily on large scales, such that fitting smaller scales mainly serves to fit the shape of nonlinearities in the redshift-space power spectrum like fingers of god (FoGs).  In fact, ref.~\cite{Chen20} found that $P_\ell$ became dominated by nonlinear, higher-order velocity statistics at around our chosen scale cut for $\ell > 0$. As an illustrative example of these effects, in Figure~\ref{fig:sig8_scan} we show the best-fit models to the $\textbf{NGCz3}$ power spectrum multipoles for a series of cosmologies with $\Omega_m$ and $h$ fixed to their Planck best-fit values and $\sigma_8$ scanned from 0.63 to 0.89.  For each $\sigma_8$ we then vary the bias parameters and counter terms to fit the $\textbf{NGCz3}$  multipoles.  Notably, the only change from varying $\sigma_8$ is in the quadrupole at $k \lesssim 0.1 \kMpc$, with the higher $k$ points having essentially the same residuals compared to the data across a broad range of power spectrum amplitudes.  The best-fitting values of the biases and counter terms that we find appear reasonable for each of the $\sigma_8$ values we tried. We can understand this effect as follows: since the linear bias is well-constrained by the monopole, varying $\sigma_8$ effectively tunes the linear quadrupole amplitude, but as effective-theory corrections like nonlinear bias turn on at higher wavenumbers this variation is erased by all the nonlinear parameters conspiring to fit a relatively smooth and featureless $P_2$. This suggests that, absent a detailed understanding of the small-scale physics underlying the bias parameters, there is not signficant information to be gained at smaller scales. Conversely, the fact that the variations in the quadrupole amplitude on large scales cannot be fit away by (reasonable) bias parameters is a demonstration that in effective theories the large scale clustering signal cannot be polluted by small-scale physics in unphysical ways.  This also suggests that future surveys capable of reducing the errors at low $k$ would significantly improve the constraints on the amplitude.

For the configuration-space side of our analysis, since we are primarily interested in fitting the BAO feature, we restrict our fitting of the post-reconstruction correlation function to $80 < s < 130\, h^{-1}$ Mpc for our fiducial setup. This range of scales effectively isolates the BAO peak and most of the \edit{additional} distance information \edit{coming from reconstruction}. It is worth noting that a fit using the same model in Fourier space would have required fitting over a wide range of wavenumbers to cover all the BAO wiggles in the power spectrum and risked numerical issues in the covariance matrix due to significant correlation with the pre-reconstruction power spectrum; the fact that the BAO feature is localized at a large scale in the correlation function\edit{, which happens to also be significantly sharpened by reconstruction at those radii,} is thus a useful fact in combining pre- and post-reconstruction data.  We also note that using broader bins in $P_\ell(k)$ could further separate the BAO and broad-band information, making the constraints even less correlated, but we have not needed to take that step for BOSS. Finally, as a consistency check we  will also want to compare fits to the pre-reconstruction correlation functions for $\textbf{z1}$ and $\textbf{z3}$ samples to our fiducial setup fitting the pre-reconstruction power spectrum; for these fits we will fit the correlation function over the entire range shown in the bottom panel of Figure~\ref{fig:data}, i.e. $r > 25 \Mpc$.

\subsection{Parameters and Priors}

We choose to sample the cosmological parameter space uniformly in $\Omega_m$, $h$ and $\ln(10^{10} A_s)$. For the purposes of our analysis we will fix the values of the baryon density $\Omega_b h^2=0.02242$ and spectral index $n_s=0.9665$ to the best-fit values from Planck \cite{Planck18-VI}, since the BOSS data are not very sensitive to these parameters which are very well-determined in the CMB and, in the case of the baryon density, also big-bang nucleosynthesis (BBN) \cite{Ivanov20}. We also fix the sum of the neutrino masses to be the minimal allowed $M_\nu = 0.06$ eV. While a number of recent works (e.g.\ ref.~\cite{Aviles20}) have sought to more exactly integrate the scale-dependent effects of massive neutrinos into LPT, in this work we will approximate the effect of neutrinos on galaxy clustering by using the ``cb'' prescription \cite{Castorina15}. This prescription is motivated by the intuition that galaxies trace the cold dark-matter and baryon fluid with linear power spectrum $P_{cb}$ and small-scale growth rate $f_c = (1 - 3 f_\nu/5) f_{\Lambda \rm CDM}$, where $f_\nu$ is the neutrino mass fraction.  This was recently shown to be an excellent approximation using phase-matched simulations in ref~\cite{Bayer21}.

For our pre-reconstruction power spectrum model we choose to fit each of the four samples with independent sets of bias parameters. As can be seen in Figure~\ref{fig:data} the NGC and SGC samples at \textbf{z1} have signficantly different power spectrum multipoles on all scales shown; indeed, using the best-fit bias parameters for a Planck cosmology for the NGC sample and convolving it with the \textbf{SGCz1} window function yields a noticeably worse fit ($\Delta \chi^2 = 19$ with our fiducial scale cuts). The multipoles at \textbf{z3} show greater agreement, and performing the same exercise of convolving the best fit model from one galactic cap with the window function of the other yields a much better fit. Nonetheless, we have opted for the more conservative approach and fit all four (statistically independent) samples pre-reconstruction using separate sets of bias parameters. Since the correlation function data were computed jointly for the NGC and SGC samples we are forced to use a unified set of bias parameters when fitting in configuration space; however, since our primary interest in fitting the post-reconstruction $\xi_\ell$'s are to extract large-scale BAO information, and since we find that the linear biases between the two galactic caps are in good agreement in our power-spectrum only fits, we do not expect this to significantly affect our results.

Our priors on the effective-theory and bias parameters are listed in Table~\ref{tab:priors}. We adopt broad, uninformative priors for the BAO broadband parameters $B_1, F$, with similarly broad priors on the polynomial broadband terms with widths such that they do not dominate the clustering signal on BAO scales. For the pre-reconstruction power spectrum we use generally broad priors for the quadratic biases $b_2, b_s$ on the assumption that they are free effective-theory coefficients of order unity. Our counterterms $\alpha_0, \alpha_2$ are set to Gaussian priors with the expectation that the galaxy power spectrum deviate from linear theory by a factor less than unity on perturbative scales. Finally, while one might expect the isotropic stochastic term $R_h^3$ to be a free parameter in the ball-park of the inverse galaxy number density $\bar{n}^{-1} \approx 3000 \Mpccube$, we put a relatively tight Gaussian prior with width $1/(3 \bar{n})$ on it given that we find an almost exact degeneracy between it and $\alpha_0$ in our fits that did not correlate significantly with any cosmological parameters for reasonable values of $R_h^3$ and $\alpha_0$. We put a physically motivated prior on $R_h^3 \sigma^2$ based on the expectation that characteristic halo velocities for BOSS LRGs are around $500$ km/s, or about $5 \Mpc$. Our pre-reconstruction fits to the pre-reconstruction correlation function follow the same set of priors as the power spectrum fits.

\begin{table}
    \centering
    \begin{tabular}{ || c || c | }
    \hline
    Parameter & Prior \\
    \hline 
    $\ln(10^{10} A_s)$ & $\mathcal{U}(1.61,3.91)$ \\
    $\Omega_m$ & $\mathcal{U}(0.20,0.40)$ \\
    $H_0$ [km/s/Mpc] & $\mathcal{U}(60.0,80.0)$ \\
    \hline
    $B_1$ & $\mathcal{U}(0,5.0)$ \\
    $F$ & $\mathcal{U}(0,5.0)$ \\
    $a_{\ell,0}$ & $\mathcal{N}(0,0.05)$ \\
    $a_{\ell,1}$ [$h^{-1}$ Mpc]  & $\mathcal{N}(0,5)$ \\
    \hline
\end{tabular}

    \begin{tabular}{ || c || c | }
    \hline
    Parameter & Prior \\
    \hline
    $(1 + b_1)\sigma_8$ & $\mathcal{U}(0.5,3.0)$ \\
    $b_2$ & $\mathcal{N}(0,10)$ \\
    $b_s$ & $\mathcal{N}(0,5)$ \\
    $\alpha_0$ [h$^{-2}$ Mpc$^2$] & $\mathcal{N}(0,100)$ \\
    $\alpha_2$ [h$^{-2}$ Mpc$^2$] & $\mathcal{N}(0,100)$ \\
    $R_h^3$ [h$^{-3}$ Mpc$^3$] & $\mathcal{N}(0,1000)$ \\
    $R_h^3 \sigma^2$ [h$^{-5}$ Mpc$^5$] & $\mathcal{N}(0,5 \times 10^4)$ \\
    \hline
\end{tabular}
    \caption{Parameter priors for our analysis. Uniform and normal distributions are indicated by $U(x_{min},x_{max})$ and $N(\mu,\sigma)$, respectively.
    }
\label{tab:priors}
\end{table}

\subsection{Test on Mocks}
\label{ssec:mock_tests}

\begin{figure}
    \begin{minipage}{.45\textwidth}
    \centering
    \includegraphics[width=\textwidth]{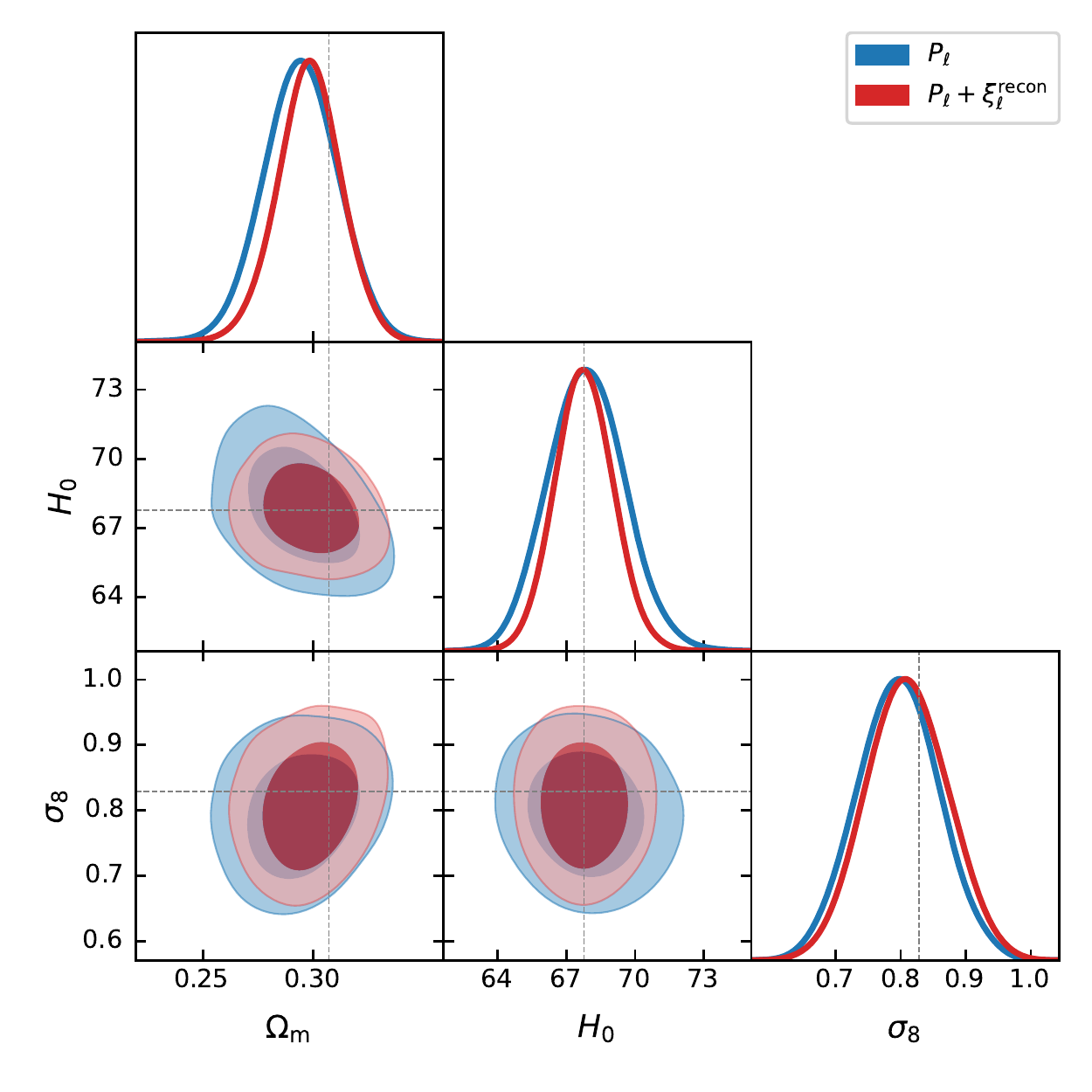}
    \end{minipage}\hfill
    \begin{minipage}{.52\textwidth}
    \centering 
    \begin{tabular}{|| c || c | c ||}
    \hline
    & $P_\ell$ & $P_\ell$ + BAO \\
    \hline 
    $\ln(10^{10} A_s)$ & $3.09\pm0.18$ & $3.10\pm0.17$  \\
    $\Omega_m$ & $0.294\pm0.017$ & $0.298\pm0.014$ \\
    $H_0$ [km/s/Mpc] & $67.9\pm1.7$ & $67.8\pm1.3$  \\
    \hline
    $\sigma_8$ & $0.797\pm0.062$ & $0.810\pm0.063$  \\
    \hline
\end{tabular}
    \end{minipage}
    
    \caption{(Left) Contours of mock constraints from the mean of 1000 Patchy mocks for the \textbf{z3} sample fitting redshift-space power spectrum monopole and quadrupole between $0.02 \kMpc < k < 0.20 \kMpc$ with (red) and without (blue) additional BAO information from the post-reconstruction correlation function near the BAO peak. Gray lines show the true cosmology of the Patchy mocks. (Right) A table of the mock constraints (mean$\pm 1\sigma$). The true cosmology is given by $\Omega_m = 0.307115,H_0 = 67.77, \sigma_8=0.8288$, well within the $1\sigma$ bounds shown.  }
    
    \label{fig:mock_results}
\end{figure}

\edit{As we have discussed, the model described in Section~\ref{sec:model}, i.e.\ LPT, has been tested extensively against simulated galaxy samples both pre- and post-reconstruction, e.g.\ in refs.~\cite{Chen21,Chen19b}. These tests were performed using periodic boxes without observational effects like window functions and realistic survey geometries; while these effects are well understood and not expected to significantly affect the accuracy of our models over the relevant scales, it is worth checking that our fiducial setup in this paper (jointly fitting the redshift-space power spectrum and post-reconstruction correlation function near the BAO peak) does not yield unexpected biases in cosmological constraints. To this end, in this subsection we apply the same analysis pipeline we will use to analyze the BOSS data to obtain mock constraints from the mean of the 1000 Patchy mocks released by the BOSS collaboration described in Section~\ref{sec:data}. In particular, we perform our test on mocks of the \textbf{z3} sample, including the NGC and SGC power spectra and combined correlation function multipoles. While these mocks employ approximate dynamics that may not exactly match the \textit{ab initio} predictions of perturbation theory, they are designed to match the survey geometry and observational systematics of the BOSS survey, and are thus a reasonable way to test whether these effects or our joint Fourier and configuration space setup meaningfully affect our results.}

\edit{The results of this test are shown in Figure~\ref{fig:mock_results}. The constraints both with and without BAO information recover the true cosmology ($\Omega_m = 0.307115,H_0 = 67.77, \sigma_8=0.8288$) of the Patchy mocks to within $1\sigma$, and adding in the post-reconstruction correlation function tightens but does not lead to significant ($<0.3\sigma$) shifts in the resulting constraints. We note that our results both with and without reconstruction show up to $0.5\sigma$ deviations from ``truth''  in both $\Omega_m$ and $\sigma_8$ despite the low statistical scatter from  averaging over 1000 mocks, which could be due to either the approximate nature of the mocks themselves or parameter projection effects; nonetheless, these results are satisfactory for our purposes since (1) they demonstrate the main goal of this subsection, which was to show that adding in the post-reconstruction correlation function does not bias our results and (2) the parameter shifts occur both with and without reconstruction, despite our having tested the latter case in simulation volumes significantly ($100\times$; \cite{Nishimichi20}) larger than the BOSS survey and with correspondingly tighter constraints and recovered unbiased constraints. Indeed, adding in post-reconstruction BAO shifts both $\Omega_m$ and $\sigma_8$ closer to ``truth'' while leaving $H_0$ firmly centered at the true value, potentially due to reduced parameter-projection effects coming from tighter constraints.}

\section{Results}
\label{sec:results}

\subsection{$\Lambda$CDM Constraints from BOSS with and without BAO}

\begin{table}[b]
    \centering
    \begin{tabular}{|| c || c | c | c ||}
    \hline
    & $P_\ell$ & $P_\ell$ + BAO & Planck \\
    \hline 
    $\ln(10^{10} A_s)$ & $\lnpk$ & $\lnjoint$ & $3.044 \pm 0.014$ \\
    $\Omega_m$ & $\Ompk$ & $\Omjoint$ &  $0.3153 \pm 0.0073$ \\
    $H_0$ [km/s/Mpc] & $\Hpk$ & $\Hjoint$ & $67.36 \pm 0.54$ \\
    \hline
    $\sigma_8$ & $\spk$ & $\sjoint$ &  $0.8111 \pm 0.0060$ \\
    \hline
\end{tabular}
    \caption{Constraints from the full BOSS sample, i.e. \textbf{NGCz1} \textbf{SGCz1}, \textbf{NGCz3} and \textbf{SGCz3}, with and without additional BAO information from the reconstructed correlation function, summarized as mean $\pm 1\sigma$. The equivalent constraints from Planck are also tabulated for comparison.}
\label{tab:main_results}
\end{table}

\begin{figure}
    \centering
    \includegraphics[width=0.45\textwidth]{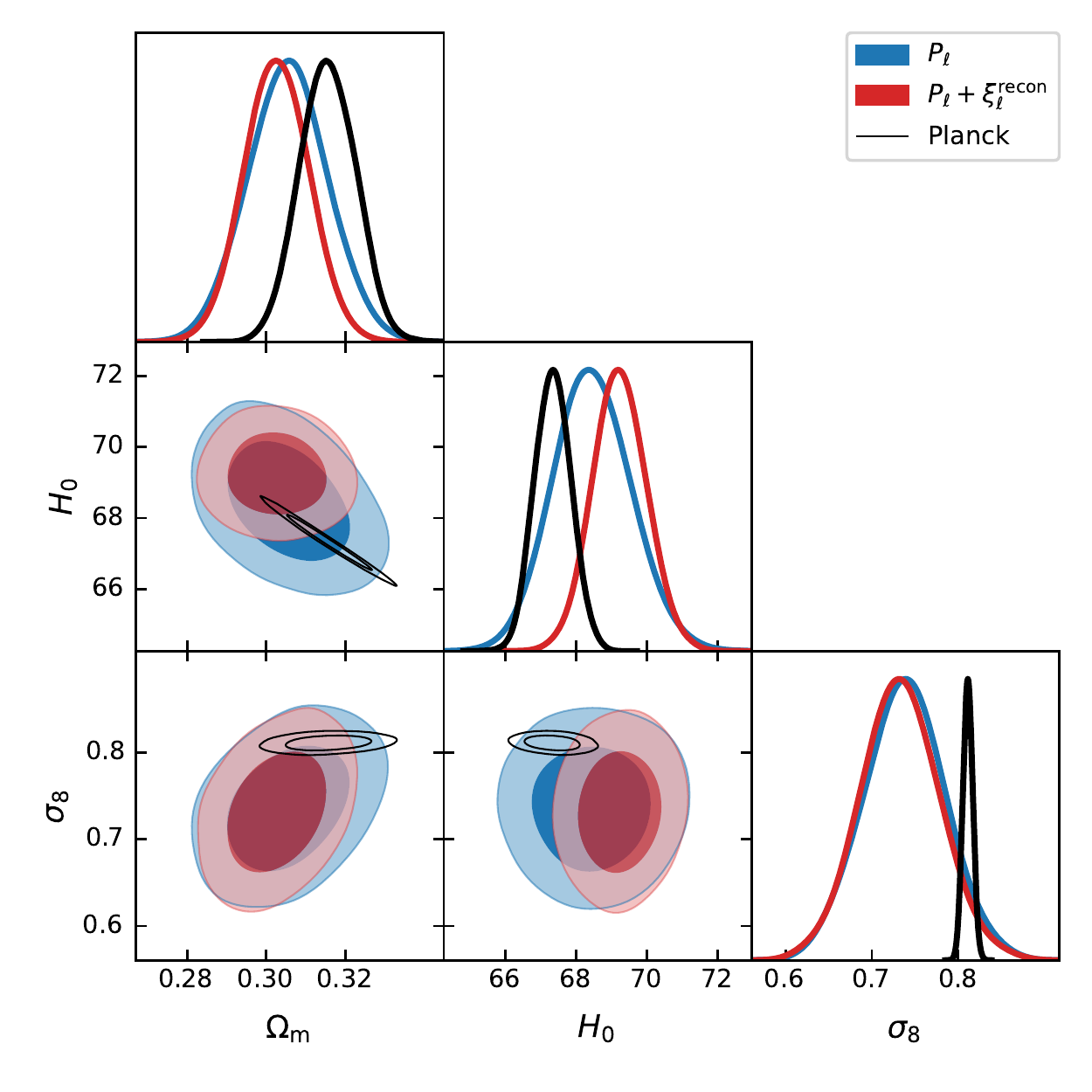}
    \includegraphics[width=0.54\textwidth]{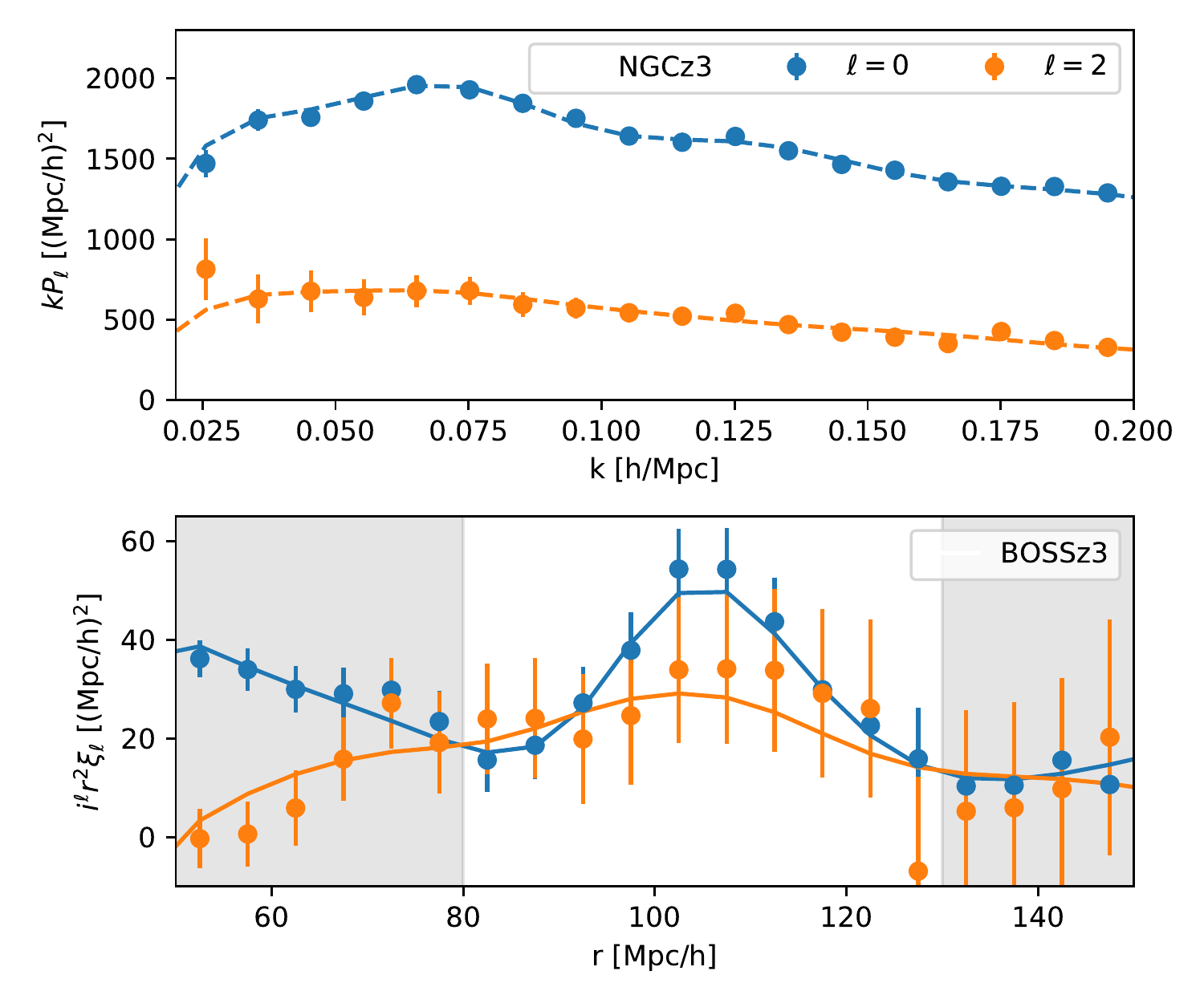}
    \caption{(Left) Constraint contours for fits to the BOSS galaxy power spectra alone (blue) and with post-reconstruction correlation function multipoles (BAO) added (red) compared to posteriors from Planck (blue), with which our constraints are broadly consistent. (Right) Binned best-fit models for the power spectrum and post-reconstruction correlation function multipoles from our chains. Here we show only the results for \textbf{NGCz3} power spectra and \textbf{z3} correlation functions for brevity; the other samples are similarly well fit, with total $\chi^2/\text{d.o.f} = 1.06.$ Gray bands in the correlation function plot show separations excluded by our fit.
    }
    \label{fig:main_contours}
\end{figure}

The main results of this paper -- constraints on $\Lambda$CDM parameters from pre-reconstruction power spectra and post-reconstruction correlation function multipoles for the full BOSS sample, including both galactic caps and redshift slices -- are shown in Figure~\ref{fig:main_contours} and listed in Table~\ref{tab:main_results}. Fitting to pre-reconstruction power spectra alone we constrain $\Omega_m = \Ompk$, $H_0=\Hpk$ and $\sigma_8=\spk$; adding in the post-reconstruction correlation function gives $\Omega_m = \Omjoint$, $H_0=\Hjoint$ and $\sigma_8=\sjoint$. When fit with a shared set of cosmological parameters, our model provides good fits to all of the individual statistics included in the likelihood (pre-recon $P_\ell$ and post-recon $\xi_\ell$ with $\ell=0$ and 2), with a combined $\chi^2/\text{d.o.f} = 1.05.$ As an example, the right panel of Figure~\ref{fig:main_contours} shows the best-fit $P_\ell$ and $\xi_\ell$ for the \textbf{NGCz3} and \textbf{z3} samples, respectively.  The other samples look similar.

Comparing the red and blue contours in Figure~\ref{fig:main_contours}, we see that the primary effect of including post-reconstruction correlation functions is to tighten constraints on the Hubble parameter $H_0$ (by around $40\%$) while also slightly tightening constraints on $\Omega_m$ and keeping the $\sigma_8$ constraint largely untouched. This is to be expected since (1) the main purpose of standard reconstruction is to sharpen the BAO peak and (2) we have fit the post-reconstruction correlation function only near the peak at $80 \Mpc < r < 130 \Mpc$. Our motivation to include the post-reconstruction correlation function in this paper was to include the information in the linear power spectrum isolated at the BAO peak in configuration space and not to use it as a further probe of nonlinearities in structure formation; indeed, the purpose of freeing the Kaiser factors $B_1$, $F$ and linear broadband polynomials in our correlation-function model was precisely to prevent systematics in reconstruction from biasing our constraints on the broadband amplitude of the linear power spectrum.

Let us conclude this subsection by comparing our results to those of other groups and experiments. As can be seen in the left panel of Figure~\ref{fig:main_contours}, our constraints from BOSS, both with and without additional information in the form of the reconstructed correlation function, are also broadly consistent with cosmic microwave background (CMB) constraints, lensing included, from Planck \cite{Planck18-VI}. Indeed, our pre-reconstruction fit has the Planck best-fit \edit{$\Omega_m$ and $H_0$} within its 1$\sigma$ contours, and $\sigma_8$ at about $1.5\sigma$ lower. \edit{This is in contrast with earlier papers using effective theory approaches to fitting the BOSS data, e.g.\ ref.'s~\cite{Ivanov20,Ivanov21,dAmico20}, which found $\sigma_8$'s significantly ($>2\sigma$) lower than Planck.} These discrepancies are potentially attributable to the data normalization issue described in Section~\ref{sec:data} as our pre-reconstruction $\Omega_m$ and $H_0$ constraints are in much better agreement with those earlier works; \edit{specifically, this normalization issue resulted in power spectra that were roughly $10\%$ too low in amplitude for a fixed window function normalization, translating to a roughly $5\%$ lower best-fit $\sigma_8$. We recover essentially identical $\sigma_8$ constraints to previous works if this correction is neglected. \edit{Indeed, after our paper was first posted to the arXiv, updated results free of window-function systematics from the authors of the aforementioned works appeared: specifically, ref.~\cite{Zhang21} used correlation functions to obtain $\sigma_8 = 0.7537^{+0.055}_{-0.06}$ and $0.7559^{+0.052}_{-0.062}$ with and without BAO, in good agreement with results from their independently measured power spectra, and ref.~\cite{Philcox21} obtained $\sigma_8 = 0.729^{+0.040}_{-0.045}$ using a window-function free power spectrum estimator and $\sigma_8 = 0.737^{+0.040}_{-0.044}$ when using the same updated BOSS power spectra as the present work, virtually identical to our pre-reconstruction constraints\footnote{For a clean comparison, we have quoted their constraints fixing the spectral tilt $n_s$ and without including the bispectrum monopole or finger-of-god reduced two-point statistics.}.} Our results are also in excellent agreement with constraints from the BOSS data using an emulator approach \cite{Kobayashi21} and the configuration-space analysis in ref.~\cite{Semenaite21}.}

Adding the post-reconstruction correlation function, which sharpens the BAO peak and tightens the $H_0$ constraint, puts the Hubble parameter best-fit from Planck at about 2$\sigma$. This is in contrast to the BOSS fits in Ref.~\cite{Philcox20}, who find $H_0 = 67.81^{+0.68}_{-0.69}$\footnote{We compare to their results keeping $n_s$ fixed rather than free for a more apples-to-apples comparison.} after adding in fits to $\tilde{\alpha}$'s from post-reconstruction power spectra despite a similar tightening of constraints, though we note that we use different post-reconstruction data ($P_\ell$ vs. $\xi_\ell$) and covariance matrices \edit{and that the mock tests in \S\ref{ssec:mock_tests} suggest that adding in the post-reconstruction correlation function does not bias our results}. Nonetheless, our constraints with and without BAO both lie on the surface of constant $\Omega_m h^3$ given by the Planck best-fit parameters; this combination is close to a principle component of the Planck posterior and is much better constrained than either $H_0$ or $\Omega_m$ alone.

\begin{figure}
    \centering
    \resizebox{0.5\textwidth}{!}{\includegraphics{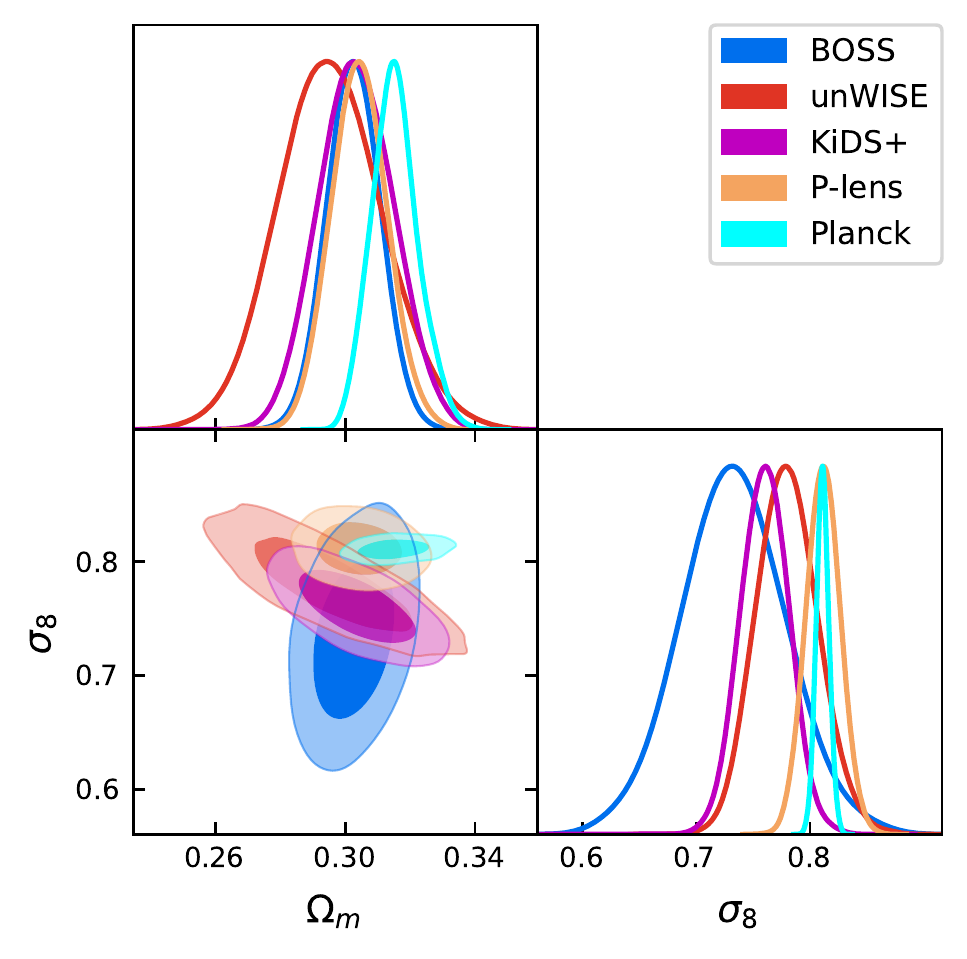}}
    \caption{A comparison of our $\sigma_8$-$\Omega_m$ constraints with a selection of other experiments, including Planck (including lensing) \cite{Planck18-VI}, Planck lensing (P-lens) with BAO prior \cite{Planck18-VIII}, KiDS+BOSS+2dFlens analysis \cite{Heymans21}, and unWISE galaxy-CMB lensing cross correlations \cite{Krolewski21}. Our BOSS constraint probes different degeneracy directions than these (primarily lensing) surveys, but is nonetheless consistent with each.
    }
    \label{fig:cmp_external}
\end{figure}

Finally, our redshift-space constraints offer an independent check to the measurements of power spectrum amplitude from weak lensing surveys like DES and KiDS, which primarily measure the hybrid quantity $S_8 = \sigma_8 (\Omega_m/0.5)^{0.5}$ and have found it to be significantly lower than the value implied by Planck ($0.832 \pm 0.013$ \cite{Planck18-VI}). Figure \ref{fig:cmp_external} summarizes these constraints in the $\Omega_m-\sigma_8$ plane. Since $\sigma_8$ and $\Omega_m$ are positively correlated in our posteriors, our constraints are not optimized to measure $S_8$; nonetheless, we obtain $S_8 = \SEightjoint$, \edit{slightly less than $2\sigma$ lower than} the Planck result and measurements of $0.775^{+0.026}_{-0.024}$ ($3\times 2$pt only) or $0.812 \pm 0.008$ (+BAO, RSD and SNIA) from DES Y3 \cite{DESY3}\footnote{We caution that these analyses, unlike ours, freed the total neutrino mass $M_\nu$.} and of $0.766^{+0.020}_{-0.014}$ from a joint KiDS, BOSS and 2dFLenS analysis \cite{Heymans21}.   At lower redshifts, once BAO and weak priors are included, the constraint from CMB lensing measured by Planck is $\Sigma_8 \equiv\sigma_8(\Omega_m/0.3)^{0.25} = 0.815\pm 0.016$ \cite{Planck18-VIII}. \edit{Our joint analysis finds} $\Sigma_8=\SigEightjoint5$, about $1.5\sigma$ lower.

\subsection{Consistency Checks}

\begin{figure}
    \centering
    \includegraphics[width=0.49\textwidth]{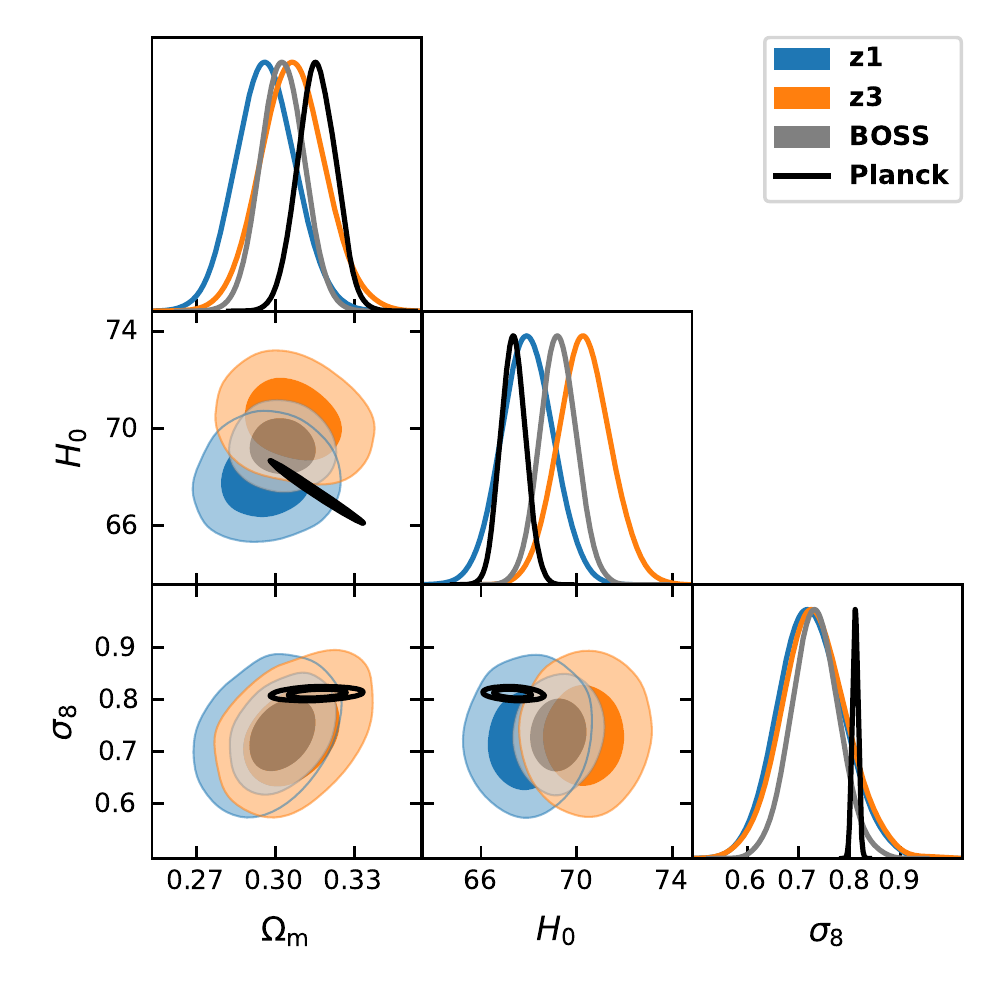}
    \includegraphics[width=0.49\textwidth]{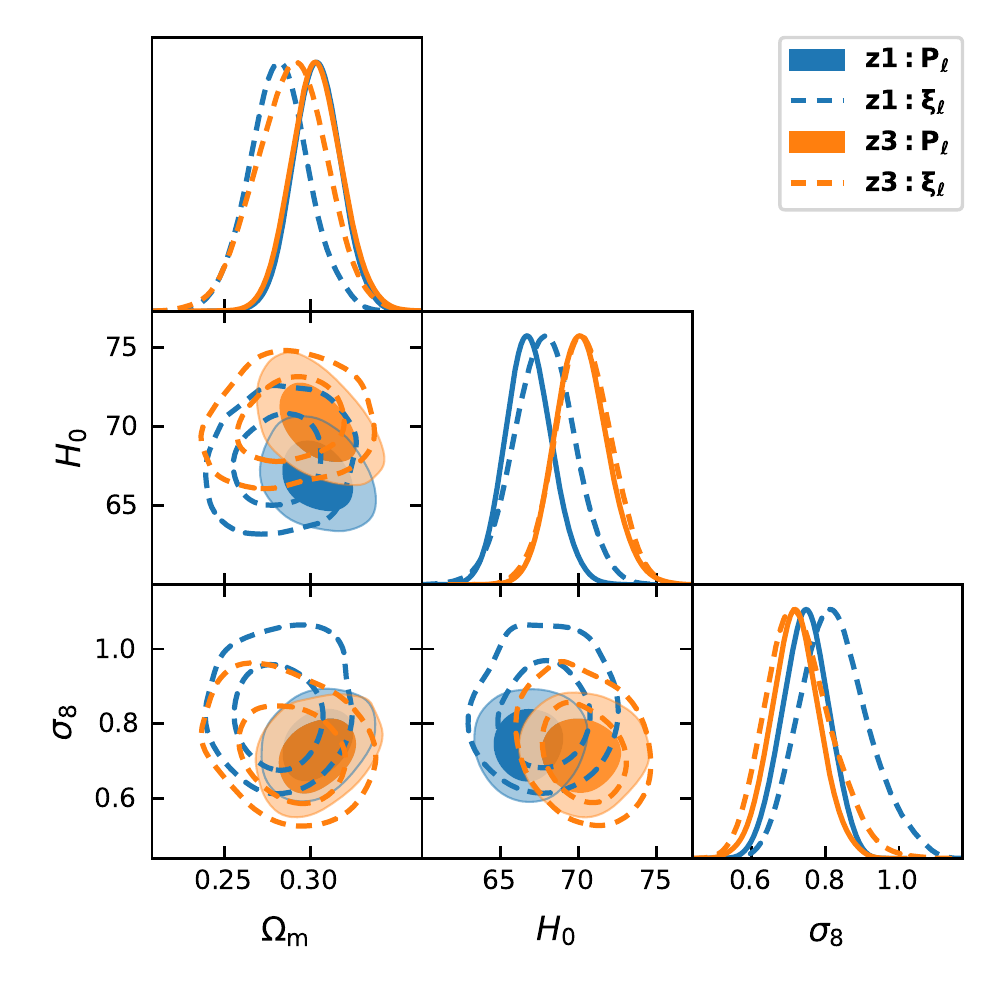}
    \caption{(Left) Constraints from the two independent redshift slices $\textbf{z1}$ (blue) and $\textbf{z3}$ (orange) compared to the joint constraint (BOSS) from both samples (gray). The two redshift bins are broadly consistent with each other, as well as with constraints from Planck (black). (Right) Constraints using pre-reconstruction power spectra and correlation functions in each of the redshift slices, fit using the same theory model.
    }
    \label{fig:consistency_contours}
\end{figure}

As discussed in Section~\ref{sec:model}, (Lagrangian) perturbation theory provides a framework within which we can model large-scale structure observables like the power spectrum and correlation function with a consistent theoretical model. This allows us to, for example, jointly model the pre-reconstruction power spectrum and post-reconstruction BAO feature without resorting to intermediate statistics like $\alpha_{\rm BAO}$'s measured from fixed-shape templates. In addition, the fact that we can model all these observables within the same framework allows us check the consistency of our model assumptions about the background cosmology and nonlinear structure formation or, alternatively, to check for systematics in each statistic. In this subsection we will describe two such consistency tests: between the high and low redshift samples and between Fourier and configuration space.

The left panel of Figure~\ref{fig:consistency_contours} shows constraints obtained by fitting the two redshift slices, \textbf{z1} and \textbf{z3}, independently while including post-reconstruction BAO information in each. The cosmological posteriors derived from the two samples are broadly consistent with each other, with significant overlap between the 1$\sigma$ regions: the $\Omega_m$ and $\sigma_8$ distributions both have means consistent to within $1\sigma$. On the other hand, while the $\bf{z1}$ sample prefers a value of $H_0$ very similar to the Planck (black contours) best fit, $\bf{z3}$ prefers values around 2$\sigma$ above Planck; we emphasize however that the combined constraints are themselves consistent with Planck. 

The right panel of Figure~\ref{fig:consistency_contours} compares fits to the pre-reconstruction power spectra (solid) and correlation functions (dashed) in each redshift slice. Again, the constraints from each $P_\ell$, $\xi_\ell$ pair are broadly consistent, with overlapping $1\sigma$ regions for all three cosmological parameters. However, while both $\sigma_8$ and $H_0$ constraints vary consistently across samples, slightly decreasing and increasing respectively with sample redshift, the $\Omega_m$ constraints are slightly lower in configuration space than Fourier space.  Given the different ranges over which we fit these statistics and the different data subsamples we do not expect perfect agreement. Nonetheless, all three parameters are broadly consistent, especially taking into account that (1) the configuration space constraints on $\Omega_m$ and $H_0$ are substantially less tight than their Fourier space counterparts and (2) unlike the power spectra, the correlation functions were computed assuming that the NGC and SGC subsamples could be meaningfully combined into one. While we find the large scale (linear) bias to be compatible between the two galactic caps at both redshifts, this does not appear to be true for the full set of bias parameters, i.e.\ taking the unconvolved best fit power spectrum from fitting the NGC and convolving it with the SGC window function does not yield a comparably good fit to the SGC power spectrum, particularly for the \textbf{z1} samples, leading to potential systematic differences between the configuration and Fourier space fits.  The differences are not important for our post-reconstruction BAO constraints, and a full re-measurement of the BOSS correlation function, and its covariance, in each galactic cap is beyond the scope of this paper.

\edit{We can also compare our results to the growth-rate ($f\sigma_8$) measurements made by the BOSS survey: our indepenent chains for each redshift bin imply $f\sigma_8(z_{\rm eff}=0.38) =0.419^{+0.035}_{-0.041}$ and  $f\sigma_8(z_{\rm eff}=0.61) = 0.422^{+0.035}_{-0.040}$, in mild tension with the official BOSS survey results.  Their consensus results (including full-shape and BAO, both using template fits) were $0.497 \pm 0.046$ and $0.436 \pm 0.035$\footnote{Here we have combined the statistical and systematic uncertainties via quadrature. A complete table of the official BOSS results can be found in Table 7 of ref.~\cite{Alam17}.}, respectively \cite{Alam17}. It is worth noting that the BOSS consensus results are a weighted combination of various analyses, including both Fourier-space and configuration-space results.  The configuration space results are not affected by issues like that of window-function normalization discussed in Section~\ref{sec:data}.  Comparison of results in each of these categories within the BOSS full-shape analysis shows differences between mean $f\sigma_8$'s greater than one standard deviation, whereas our analysis using the new window function normalizations yield Fourier and configuration space means within $1\sigma$; for \textbf{z1} we get $f\sigma_8$ constraints of $0.434\pm 0.038$ and $0.470 \pm 0.054$ and for \textbf{z3} we get $0.413\pm 0.039$ and $0.414 \pm 0.050$ in Fourier and configuration space, respectively, pre-reconstruction.}

\section{Conclusions}
\label{sec:conclusions}

Galaxy redshift surveys are an important source of cosmological information, allowing us to constrain properties of the early universe and general relativity through measurements of redshift-space distortions and baryon acoustic oscillations in galaxy clustering. Recent developments in cosmological perturbation theory, particularly in the arena of effective theories and IR resummation, have further put these data on rigorous and precise footing, making it possible to consistently fit a range of measurements from these surveys within a consistent framework. In the next few years, surveys like DESI \cite{DESI} and Euclid \cite{EUCLID} will probe larger volumes at higher redshifts, greatly increasing our constraining power on cosmological parameters while making accurate modeling on quasilinear scales ever more important.

In this paper we have presented an analysis of the pre- and post-reconstruction power spectra and correlation functions from the BOSS survey \cite{Dawson13} within the framework of Lagrangian perturbation theory. Unlike previous works, we do not combine pre- and post-reconstruction data through additional fitting parameters for the BAO (e.g. $\tilde{\alpha}_{\parallel,\perp}$) whose covariances are determined from mocks. Rather, for a given set of cosmological parameters ($\Omega_m, h, \sigma_8$) and galaxy bias coefficients we compute directly the power spectrum and correlation function as predicted by perturbation theory and compare them with observations to compute the likelihood.

For our main result, we jointly fit pre-reconstruction power spectrum and post-reconstruction correlation function multipoles for the full BOSS sample. In order to avoid undue correlation with the power spectrum measurements, we take advantage of the fact that the BAO signal is well-isolated in the correlation function, particularly after reconstruction, and only fit the correlation function in a narrow band containing the peak ($80 \Mpc < r < 130 \Mpc$). Our results are consistent with constraints from Planck, as well as with $S_8$ measurements from weak lensing surveys. We have further checked our analyses by considering constraints from each of the redshift slices ($z_{\rm eff} = 0.38$, 0.61) with and without post-recon BAO, and in the former case by fitting both the power spectrum and correlation function, finding that constraints from each subsample or observable are broadly consistent.

Let us conclude by pointing out some possible future directions. A natural extension of this work is to include further observables in our analysis: LPT in the context of galaxy-lensing cross correlations has been studied in ref.~\cite{Modi17,Wilson19,Kitanidis21} and formed the basis of the model applied to the unWISE data in ref.~\cite{Krolewski21}, as well as luminous red galaxies (LRGs) from DESI in ref.~\cite{White21}. Recent work \cite{Modi20,Kokron21,Hadzhiyska21,Zennaro21} combining the Lagrangian bias scheme used in this paper with nonlinear dynamics from N-body simulations should further extend the reach and applicability of LPT for analyzing lensing cross correlations within a Lagrangian framework.  This would allow for a consistent analysis of lensing surveys along with RSD and BAO (as shown in this paper) which will become very powerful in the era of DESI \cite{DESI}, Euclid \cite{EUCLID}, Rubin \cite{LSST} and CMB-S4 \cite{CMBS4}. In parallel, our analysis can be extended to include other physical effects such as relative baryon-dark matter perturbations or more exotic early-universe physics such as early dark energy \edit{light relics} or primordial features in the power spectrum; the modeling of these effects have been studied within LPT \cite{Chen19,Chen20b,Rampf21} and applied to data within Eulerian perturbation theory without reconstruction \cite{Beutler17b,Schmidt17,Ivanov20b,DAmico21EDE}. \edit{Many of these signatures are sharpened by reconstruction.} Another potential effect is that of anisotropic secondary bias due to line-of-sight selection biases, which have the potential to skew measurements of $\sigma_8$ from RSD \cite{Hirata09}. These effects have so far not been modeled with LPT, though the equivalent Eulerian framework have been explored in e.g.\ Ref.~\cite{Desjacques18}. We discuss the current status of evidence for these effects in Appendix~\ref{app:anis}. Of course, many of these additional effects will likely be better constrained by combining data: for example, relative baryon-dark matter perturbations have a potential to bias BAO measurements but conversely, by including its effects in a theory model, it may be easier to constrain the size of their effect on galaxy clustering when post-reconstruction data is included in the analysis. We leave these developments for future work.

\acknowledgments
We thank Pat McDonald for useful discussions about, and help with, the window functions used in this work. We thank Mariana Vargas Maga\~{n}a for providing data and mock measurements of the BOSS correlation functions.
S.C.\ is supported by the National Science Foundation Graduate Research Fellowship (Grant No.~DGE 1106400) and by the UC Berkeley Theoretical Astrophysics Center Astronomy and Astrophysics Graduate Fellowship. M.W.~is supported by the U.S.~Department of Energy and by NSF grant number 1713791. Z.V. acknowledges the support by the Kavli Foundation.
We acknowledge the use of \texttt{Cobaya} \cite{CobayaSoftware,Cobaya}, \texttt{Class} \cite{CLASS}, \texttt{GetDist} \cite{GetDist} and \texttt{velocileptors} \cite{Chen20} and thank their authors for making these products public.  This research used resources of the National Energy Research Scientific Computing Center (NERSC), a U.S.\ Department of Energy Office of Science User Facility operated under Contract No.\ DE-AC02-05CH11231.
This work made extensive use of the NASA Astrophysics Data System and of the {\tt astro-ph} preprint archive at {\tt arXiv.org}.

%======================================================%
\appendix
%======================================================%
%======================================================%

\section{Fast Evaluation via Taylor Series}
\label{app:taylor}

\begin{figure}
    \centering
    \includegraphics[width=\textwidth]{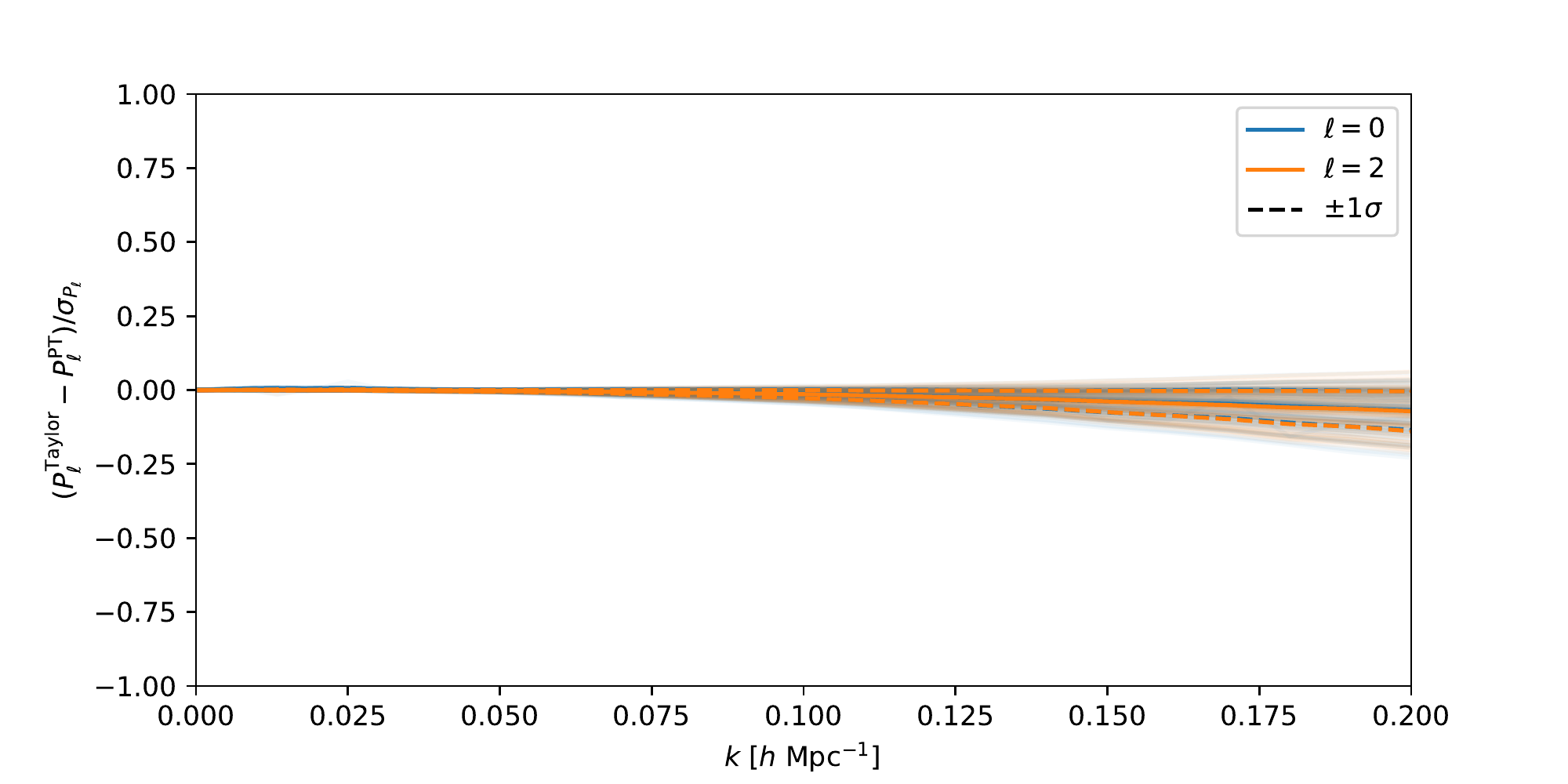}
    \caption{Differences between unconvolved power spectra computed directly from \texttt{CLASS} and \texttt{velocileptors} vs. approximated by a 4$^{\rm th}$-order Taylor series, sampled from a chain fitting the $\textbf{NGCz3}$ power spectrum, compared to the data error bars on that sample. Solid lines show mean deviation while $\pm 1\sigma$ deviations are shown as dashed lines. The monopole and quadrupole deviations are shown in blue and orange, respectively.
    }
    \label{fig:taylor}
\end{figure}

In order to speed up our model evaluations (obtaining transfer functions from \texttt{CAMB} alone takes a few seconds per cosmology) we use a Taylor series centered around a representative cosmology ($\Omega_m,\, h, \, \sigma_8 =  0.31, 0.68, 0.73$\footnote{This central $\sigma_8$ value was chosen prior to discovering the window-function normalization issue resulting in systematically low $\sigma_8$ constraints. The resulting Taylor-series predictions are nonetheless sufficiently accurate for our final constraints, as shown below. }) to evaluate the perturbation-theory predictions of each cosmology in our chains. This technique has previously been applied to both full-shape forecasts \cite{Mergulhao21} and data analyses of BOSS data \cite{dAmico20,Colas20}, and even at the linear level to approximate the predictions of a hybrid N-body/Lagrangian bias model of real-space clustering \cite{Hadzhiyska21}. Specifically, both the pre-reconstruction power spectrum and post-reconstruction correlation functions (minus broadband terms) can written as inhomogeneous quadratic polynomials in the bias parameters, e.g. for a fixed cosmology $\Theta$
\begin{equation*}
    \big( P_0(\bk), P_2(\bk), P_4(\bk), \xi_0(\bs), \xi_2(\bs) \big)(\Theta) = \beta_i \beta_j \mathcal{M}^{ij}(\Theta)
\end{equation*}
where the vector $\beta = (1, b_1, b_2, b_s, \alpha_0, \alpha_2, R_h^3,  R_h^3 \sigma^2, B_1, F)$. Each component of the matrix $\mathcal{M}^{ij}$ is a smooth function of the cosmological parameters so can be approximated as
\begin{equation*}
    \mathcal{M}^{ij}(\Theta) = \sum_{n=0}^N \frac{1}{n!}\,  (\Theta - \Theta_0)_{i_1} ... (\Theta - \Theta_0)_{i_n}\, \partial_{i_1 ... i_n } \mathcal{M}^{ij}(\Theta_0).
\end{equation*}
We evaluate the derivatives numerically using the publicly available \texttt{FinDiff}\footnote{ \url{https://findiff.readthedocs.io/en/latest/} } package. In order to take advantage of central differences these derivatives are computed using a grid with $2N + 1$ points along each axis, where we find $N = 4$ to be sufficient for any cosmology relevant for constraints from any of the sets of samples considered in our analysis. In Figure \ref{fig:taylor} we show differences between the unconvolved power spectrum multipoles as computed directly from \texttt{velocileptors} and the Taylor series approximation for 100 representative elements of a power-spectrum fit to $\textbf{NGCz3}$, compared to the much larger error bars of the data itself. Producing the grids used to compute derivatives using one core on \texttt{Cori}\footnote{\url{https://docs.nersc.gov/systems/cori/}} takes about three minutes, while evaluating the Taylor series itself given the coefficients takes less than a hundredth of a second.

\section{Nonlinear Damping of the BAO within RecIso}
\label{app:sigma_bao}
In this appendix we give the specific form of the BAO damping in the reconstructed power spectrum within the \textbf{RecIso} scheme. The reconstructed field is defined to be the difference between the overdensities of \textit{displaced} galaxies $d$ and \textit{shifted} galaxies $s$, i.e. $\delta_{\rm recon} = \delta_d - \delta_s$. Theoretical modeling of the nonlinear damping of the reconstructed field involves calculating cross correlations between the Lagrangian displacements of the two fields. Thus in general we expect each piece of
\begin{equation*}
    P^{\rm recon}_{s}(k,\mu) = P^{dd}_s - 2 P^{ds}_s + P^{ss}_s, \quad P^{ab}_s = P^{ab}_{lin, nw} + e^{-\half k^2 \Sigma^2_{ab}(\mu)} P^{ab}_{lin, w} + ..., 
\end{equation*}
to have a different damping form, where $P^{ab}_{lin}$ refers to the (undamped) linear-theory predictions for each spectrum and $\Sigma^2_{ab}$ its damping parameter. Specifically, $\Sigma^2_{ab}$ is the isotropic component of the displacement two-point function
\begin{equation}
    A^{ab}_{ij} = \avg{\Delta^{ab}_i \Delta^{ab}_j}, \; \Delta^{ab} = \Psi^a(\bq_1) - \Psi^b(\bq_2).
\end{equation}
evaluated at the BAO scale, i.e. $\Sigma^2_{ab} = \third \delta_{ij} A^{ab}_{ij}$ for $q = r_d$.

Applying the above logic we then have the linear-theory forms
\begin{align*}
    P^{dd}_{lin}(k,\mu) &= \Big(b - \mathcal{S} + f\mu^2 (1 - \mathcal{S}) \Big)^2 P_{\rm lin}(k) \nonumber \\
    P^{ds}_{lin}(k,\mu) &= - \mathcal{S}\, \Big(b - \mathcal{S} + f\mu^2 (1 - \mathcal{S}) \Big) P_{\rm lin}(k) \nonumber \\
    P^{ss}_{lin}(k,\mu) &= \mathcal{S}^2 P_{\rm lin}(k)
\end{align*}
and damping parameters
\begin{align*}
    \Sigma^2_{dd}(\mu) &= (1 + f(2+f)\mu^2)\, \Big( 2\tilde{\Sigma}^2_{dd}(0)  - 2 \tilde{\Sigma}^2_{dd}(r_d) \Big) \\
    \Sigma^2_{ds}(\mu) &= (1 + f(2+f)\mu^2) \tilde{\Sigma}^2_{dd}(0) + \tilde{\Sigma}^2_{ss}(0) - 2 (1 + f\mu^2) \tilde{\Sigma}^2_{ds}(r_d) \\
    \Sigma^2_{ss}(\mu) &=  2\tilde{\Sigma}^2_{ss}(0)  - 2 \tilde{\Sigma}^2_{ss}(r_d) 
\end{align*}
where the tilded $\Sigma$'s are defined as
\begin{align}
    \tilde{\Sigma}_{dd}^2(r) &= \frac{1}{3} \int \frac{dk}{2\pi^2} \Big[\big( 1 - \mathcal{S}\big)^2\, P_{\rm lin}(k)\ j_0(k r) \Big] \nonumber \\
    \tilde{\Sigma}_{ss}^2(r) &= \frac{1}{3} \int \frac{dk}{2\pi^2} \Big[  \mathcal{S}^2\, P_{\rm lin}(k)\ j_0(k r) \Big] \nonumber \\
     \tilde{\Sigma}_{ds}^2(r) &= \frac{1}{3} \int \frac{dk}{2\pi^2} \Big[ -\mathcal{S}\, \big( 1 - \mathcal{S}\big)\, P_{\rm lin}(k)\ j_0(k r) \Big].
\end{align}
We refer the readers to \cite{Chen19b} for a detailed derivation of these damping parameters.

%======================================================%
\section{Parameters for BAO Fit}
\label{app:template_fit}

Our goal in this appendix is to explicitly spell out the assumptions of the standard template fits to the BAO (e.g. in \cite{Beutler17,Vargas18,Philcox20,DAmico21}). We begin by writing down Alcock-Paczynski scalings needed to convert between fiducial and true cosmology before writing down the additional assumptions that result in the traditional template fit with $r_d$-dependent BAO parameters.

\subsection{Alcock-Paczynski Effect}

Galaxy survey data are typically presented in units wherein redshifts and angles are translated into distances via the Hubble function $E^{\rm fid}(z)$ and angular-diameter distance $D^{\rm fid}_A(z)$ of some fiducial cosmology\footnote{For concreteness we will assume throughout that all distances are given in $h^{-1}$ units.}. Given this choice of coordinates it is necessary given any cosmological model $\Theta$ to rescale theory predictions from the true physical separations in the theory into theory units. For the anisotropic power spectrum this is given by
\begin{equation}
    P^\obs_s(k^{\obs}_\parallel,k^{\obs}_\perp) = \alpha_{\parallel}^{-1} \alpha_\perp^{-2}  \, P^\true(k^\true_\parallel,k^\true_\perp), \quad k^\true_{\parallel,\perp} = \frac{k^\obs}{\alpha_{\parallel,\perp}}
\end{equation}
where the Alcock-Paczynski (AP) parameters are defined as
\begin{equation}
    \alpha_\parallel = \frac{E^\fid(z)}{E(z)}, \quad \alpha_\perp = \frac{D_A(z)}{D^\fid_A(z)}.
\end{equation}
For covenience we will drop the superscript for the ``true'' quantities in the remainder of this appendix. In terms of these quantities we can write the true wavenumber magnitude and angle as \cite{Beutler17}
\begin{align}
    k &= \alpha_\perp^{-1} \sqrt{1 + \mu_\obs^2 (F_{\rm AP}^{-2} - 1)}\, k^\obs \nonumber \\
    \mu  &= \frac{1}{F_{\rm AP}} \Big( \sqrt{1 + \mu_\obs^2 (F_{\rm AP}^{-2} - 1)}\Big)^{-1}\, \mu^\obs
\end{align}
where $F_{\rm AP} = \alpha_\parallel/\alpha_\perp$. Note that both $F_{\rm AP}$ and the $\mu$ are invariant under isotropic scaling $\alpha_{\parallel,\perp} \rightarrow c\alpha_{\parallel,\perp}$. 

\subsection{BAO in the Power Spectrum}

In standard template fits to the BAO the wiggles from a fixed ``template'' power spectrum are rescaled to model the observed BAO signal. In this case we must allow for the possibility that the true linear power spectrum of the universe has BAO wiggles with a slightly different shape.

The trick is to assume that the linear power spectrum for a given cosmology $\Theta$ can be split into a smooth ``no-wiggle'' component $P_{nw}$ and a ``wiggle'' component $P_w$ whose cosmology dependence can be approximated in terms of its amplitude $A$ and a template scaled by the BAO radius $r_d$:
\begin{equation}
    P_{\rm lin}(k) = P_{nw}(k|\Theta) +  A(\Theta) g(r_d(\Theta) k).
\label{eqn:Plin-Adefn}
\end{equation}
In particular, for a given ``template'' power spectrum at the fiducial cosmology we can extract the BAO template
\begin{equation}
    G^\fid(k) = A^\fid g(r^\fid_d k).
\end{equation}
The ``wiggle'' component should be understood to be unique only up to a smooth (polynomial) broadband that does not carry scale information.

The anisotropic redshift-space power spectrum, taking into nonlinear BAO damping due to bulk displacements, is
\begin{equation*}
    P_s(k,\mu) = (b + f\mu^2)^2 \big( P_{nw}(k) + e^{-\half k^2 \Sigma^2(\mu)} P_w(k) \big) + ...
\end{equation*}
where the ellipsis stands for higher order terms, which we will assume can be absorbed by smooth broadband terms, particularly post-reconstruction. Combining with the above parametrization of the linear power spectrum we get that the observed linear power spectrum is
\begin{equation}
    P^\obs_s(k_\obs,\mu_\obs) = \alpha_\parallel^{-1} \alpha_\perp^{-2} (b + f\mu^2)^2 \Big( P_{nw}(k) + e^{-\half k^2 \Sigma^2(\mu)} A(\Theta) g(r_d k) \Big) + ...,
\end{equation}
In particular, the damped ``wiggle'' component takes the form
\begin{align*}
    P^\obs_{w} &= \alpha_\parallel^{-1} \alpha_\perp^{-2} (b + f\mu^2)^2 e^{-\half k^2 \Sigma^2(\mu)} A g(r_d k) \\
    &= \alpha_\parallel^{-1} \alpha_\perp^{-2} (b + f\mu^2)^2 e^{-\half k^2 \Sigma^2(\mu)} A\, g\Bigg(r^\fid_d \Big(\frac{r_d}{r^\fid_d}\Big)  \alpha_\perp^{-1} \sqrt{1 + \mu_\obs^2 (F_{\rm AP}^{-2} - 1)} k_\obs \Bigg) \\
    &= \Bigg(\frac{r^\fid_d}{r_d}\Bigg)^{-3}\, \tilde{\alpha}_\parallel^{-1} \tilde{\alpha}_\perp^{-2} (b + f\mu^2)^2 e^{-\half k^2 \Sigma^2(\mu)} \Big(\frac{A}{A^\fid}\Big)\, G^\fid\Bigg( \tilde{\alpha}_\perp^{-1} \sqrt{1 + \mu_\obs^2 (\tilde{F}_{\rm AP}^{-2} - 1)}\, k_\obs\Bigg) \\
    &\equiv \tilde{A} (b + f\mu^2)^2 e^{-\half k^2 \Sigma^2(\mu)} G^\fid\Bigg( \tilde{\alpha}_\perp^{-1} \sqrt{1 + \mu_\obs^2 (\tilde{F}_{\rm AP}^{-2} - 1)}\, k_\obs\Bigg) \\
    &\equiv (B + F\mu^2)^2 e^{-\half k^2 \Sigma^2(\mu)} G^\fid\Bigg( \tilde{\alpha}_\perp^{-1} \sqrt{1 + \mu_\obs^2 (\tilde{F}_{\rm AP}^{-2} - 1)}\, k_\obs\Bigg)
\end{align*}
where we have dropped the cosmology dependence in $A = A(\Theta)$ and $r_d = r_d(\Theta)$ and defined the modified AP parameters
\begin{equation}
    \tilde{\alpha}_{\parallel,\perp} = \Bigg(\frac{r^\fid_d}{r_d}\Bigg)\ \alpha_{\parallel,\perp},\quad \tilde{F}_{\rm AP} \equiv \frac{\tilde{\alpha}_\parallel}{\tilde{\alpha}_\perp} = F_{\rm AP}
\end{equation}
casting the BAO measurement specifically as one of ratios of cosmological distances with the sound horizon. Note that the ratio $\beta = F/B = f/b$ is equal to that in the normal Kaiser formula and is invariant to the AP effect and template normalization. For the sake of brevity we will not repeat the above derivation for the case of reconstruction but note that the essential features (i.e. $r_d$ scaling with fixed $F_{\rm AP}$) are unchanged in that case.

In practice we would like to expand the power spectrum, broadband included, about the fiducial cosmology, at which all the ratios $X/X^\fid$ are equal to unity. In this case, the ``wiggle'' component can be exactly rescaled as above while changes in the broadband power spectrum are expected to be smooth and degenerate with smooth polynomials, i.e.
\begin{align}
    P^\obs_s(k_\obs,\mu_\obs) &= (B + F\mu_\obs^2)^2 P^\fid_{nw}(k_\obs)  + (B + F\mu^2)^2 \nonumber \\
    &e^{-\half k^2 \Sigma^2(\mu)} G^\fid\Bigg( \tilde{\alpha}_\perp^{-1} \sqrt{1 + \mu_\obs^2 (\tilde{F}_{\rm AP}^{-2} - 1)}\, k_\obs\Bigg)  + \sum_{n,m}\, a_{nm} k^n \mu^{2m}.
\end{align}
where $P^\fid_{nw}$ is the ``no-wiggle'' power spectrum at the fiducial cosmology. In practice we work with multipoles, in which case we can write
\begin{equation}
    P^\obs_\ell(k_\obs) = \Big( ... \Big)_\ell + \sum_{n}\, a_{\ell,n} k^n
\end{equation}
where the ellipses stand for multipoles of the non-polynomial terms. Equivalently, for configuration space analyses we can simply supplement the Fourier transform of these terms with a polynomial in the (inverse) radius:
\begin{equation}
    \xi^\obs_\ell(s_\obs) = {\rm FT}\left\{ (...)\right\}_\ell + \sum_n b_{\ell,n} s_\obs^{-n}
\end{equation}
where we have implicitly used that the Fourier transform of the smooth broadband is also smooth over the range of interest.

\section{Anisotropic secondary bias}
\label{app:anis}

Our analysis, in common with most other analyses in the field, has assumed that the probability that a galaxy makes it into the sample is independent of the large-scale tidal field in which that galaxy sits.  Specifically, we assume that the galaxy overdensity is a function of scalar quantities that can be constructed from second and higher derivatives of the gravitational potential.  Since several ``non-scalar'' halo properties, like shapes and angular momenta, depend upon the large-scale tidal fields in which the halos are situated \cite{Obuljen19} there is the possibility that important galaxy properties also inherit this dependence and if the probability of the galaxy appearing in the catalog with a successful redshift depends upon those properties we would introduce a non-scalar component to the bias \cite{Hirata09,Desjacques18}.  This would invalidate our analysis.

The case for or against anisotropic secondary (or ``assembly'') bias for BOSS galaxies is currently uncertain \cite{Martens18,Obuljen20,Singh21}.  The strongest claim so far is that of ref.~\cite{Obuljen20}, who argue they have detected such a bias at $5\,\sigma$ by splitting the galaxies using a combination of stellar mass and line-of-sight velocity dispersion.  Specifically they show that for two subsamples of galaxies, selected in the $M_\star-\sigma_\star$ plane, they can obtain different power spectrum quadrupoles while matching the power spectrum monopole.  Within linear theory and in the absence of anisotropic bias $P(k,\mu)=(b+f\mu^2)^2P_{\rm lin}(k)$ \cite{Kaiser87} and so matching the monopoles should imply that the quadrupoles also match.  However in the presence of anisotropic secondary bias the prefactor becomes $(b+f\mu^2+b_q[\mu^2-1/3])^2$, where $b_q$ is the anisotropic bias \cite{Obuljen20} and it is possible to have different quadrupoles while matching the monopoles.

We should treat this claim with caution, since it is based upon a very simple model.  Indeed, we have already seen that samples of galaxies in the NGC and SGC regions of the BOSS survey can have quite similar monopoles with relatively different quadrupoles.  A careful look at Figs.~3-6 of ref.~\cite{Obuljen20} shows that most of the evidence for anisotropic secondary bias arises from $k>0.1\,h\,{\rm Mpc}^{-1}$ where the linear analysis for the quadrupole is completely inadequate.  As one example, our models contain terms going as $k^2\,P_{\rm lin}(k)$ that can have different amplitudes between the monopole and quadrupole for different samples, showing that the high-$k$ part of the quadrupole can be quite different than linear theory estimates may imply.  Phrased differently, models that vary significantly in their satellite content and finger-of-god can cause variations in the quadrupole at high $k$ that are significantly larger than the variations in the monopole that they induce.

\begin{figure}
    \centering
    \resizebox{\columnwidth}{!}{\includegraphics{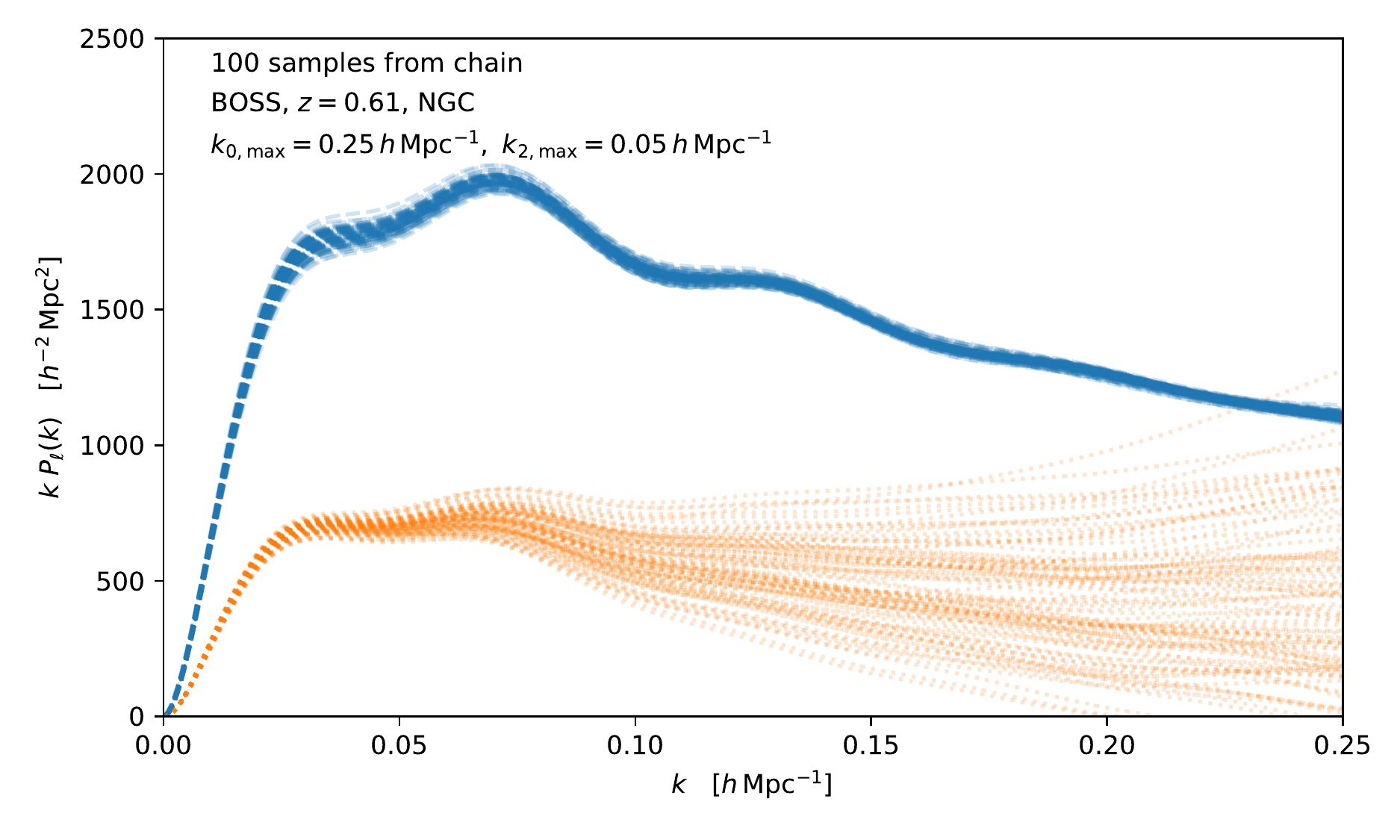}}
    \caption{Model power spectra for 100 models chosen at random from a Markov chain fit to the $z=0.61$ NGC power spectrum with $k_{\rm max}=0.25\,h\,\mathrm{Mpc}^{-1}$ for the monopole but $k_{\rm max}=0.05\,h\,\mathrm{Mpc}^{-1}$ for the quadrupole.  All of the models are thus consistent within statistical errors with a constant monopole while being essentially unconstrained as to the quadrupole.  Note the monopoles (blue dashed lines) form a tight envelope while the quadrupoles (orange dotted lines) agree only at low $k$.  By $k\simeq 0.1\,h\,\mathrm{Mpc}^{-1}$ the differences are about a factor of $2$.  }
    \label{fig:fixed_mono_only}
\end{figure}

To get a sense for how much non-linear bias and non-linear dynamics within the standard model (without anisotropic secondary bias) can modify the quadrupole at fixed monopole we did the following experiment.  We picked the $z=0.61$ NGC sample, and fit our standard pre-reconstruction power spectrum model to $k\simeq 0.25\,h\,\mathrm{Mpc}^{-1}$ for the monopole but only $k\simeq 0.05\,h\,\mathrm{Mpc}^{-1}$ for the quadrupole\footnote{Since the quadrupole is now highly unconstrained we narrowed the prior on $\alpha_2$ to be $\mathcal{N}(0,25)$, corresponding to a finger-of-god velocity dispersion of $\mathcal{O}(500\mathrm{km}\,\mathrm{s}^{-1})$ which is comparable to the values of $\alpha_2$ we obtain from the fits in the main body of the paper.} at fixed cosmology.  We then sampled 100 models from the chain, each model having equal monopoles within statistical errors, to see how much the quadrupoles can differ as a function of $k$.  The results are shown in Fig.~\ref{fig:fixed_mono_only}.    Note the monopoles (blue dashed lines) form a tight envelope while the quadrupoles (orange dotted lines) agree only at low $k$.  By $k\simeq 0.1\,h\,\mathrm{Mpc}^{-1}$ the differences are about a factor of $2$. 

Based upon this calculation we believe the case for anisotropic secondary bias in the BOSS galaxies remains unproven.  While our calculations do not prove the absence of such an effect, existing measurements are also consistent with differences expected in currently popular models that neglect these effects.  Given its importance as a source of systematic errors for future redshift surveys, further investigation is clearly warranted.

%======================================================%

\bibliographystyle{JHEP}
\bibliography{main}
\end{document}